# Measuring Oxygen, Carbon Monoxide and Hydrogen Sulfide Diffusion Coefficient and Solubility in Nafion Membranes

**Vijay A. Sethuraman,[a] Saahir Khan,[b, c] Jesse S. Jur[d], Andrew T. Haug,[e] and John W. Weidner[*]**

Center for Electrochemical Engineering, Department of Chemical Engineering
University of South Carolina, Columbia, South Carolina 29208, USA

A Devanathan-Stachurski type diffusion cell made from a fuel cell assembly is designed to evaluate the gas transport properties of a proton exchange membrane as a function of cell temperature and gas pressure. Data obtained on this cell using the electrochemical monitoring technique (EMT) is used to estimate solubility and diffusion coefficient of oxygen ($O_2$), carbon monoxide (CO) and hydrogen sulfide ($H_2S$) in Nafion membranes. Membrane swelling and reverse-gas diffusion due to water flux are accounted for in the parameter estimation procedure. Permeability of all three gases was found to increase with temperature. The estimated activation energies for $O_2$, CO and $H_2S$ diffusion in Nafion 112 are 12.58, 20 and 8.85 kJ mol$^{-1}$, respectively. The estimated enthalpies of mixing for $O_2$, CO and $H_2S$ in Nafion 112 are 5.88, 3.74 and 7.61 kJ mol$^{-1}$, respectively. An extensive comparison of transport properties estimated in this study to those reported in the literature suggests good agreement. Oxygen permeability in Nafion 117 was measured as a function of gas pressures between 1 and 3 atm. Oxygen diffusion coefficient in Nafion 117 is invariant with pressure and the solubility increases with pressure and obeys Henry's law. The estimated Henry's constant is 3.5 x $10^3$ atm.

Keywords: Carbon monoxide, Hydrogen sulfide, Oxygen, Diffusion, Solubility, Nafion, Electrochemical monitoring technique.

---

[a] – Present address: Environmental Energy Technology Division, Lawrence Berkeley National Laboratory, Berkeley, CA 94720, USA.
[b] – NSF-REU participant from Stanford University, Stanford, CA 94305, USA.
[c] - Present address: Feinberg School of Medicine, Northwestern University, Evanston, IL 60208, USA.
[d] – Present address: Department of Chemical Engineering, North Carolina State University, Raleigh, NC 27695, USA.
[e] – Present address: 3M Center, 3M Fuel Center Components, St. Paul, Minnesota 55144, USA.
[*] – Corresponding author address: Professor, Department of Chemical Engineering, 3C05 Swearingen Engineering Center, University of South Carolina, 315 Main Street, Columbia, South Carolina 29208, USA.
Phone: + 1 (803) 777-3207; Fax: +1 (803) 777-8265; E-mail: weidner@cec.sc.edu

## 1. INTRODUCTION

Gas diffusion across a proton exchange membrane (PEM) of a PEM fuel cell has important consequences for its performance and its durability [1,2]. Very thin membranes (~ 25 μm and less) are presently used in PEM fuel cells in order to decrease the membrane resistance and to increase power density. However, the direct negative consequence of using thin membranes is the increased crossover of reactants from one side to the other. In addition to this loss of fuel due to crossover, recent studies on PEM fuel cell durability suggest that $H_2$ and $O_2$ crossover play an important role in the reduction of Pt ions (diffusing from the cathode) inside the membrane [3, 4]. The diffusing gases further fuel the hydroxyl ($HO^•$) and hydroxyl-peroxyl ($HOO^•$) formation reactions on metallic catalyst particles present inside the membrane. These radicals are known to attack the tertiary hydrogen at the α-carbon of the perfluorinated membranes commonly used in PEM fuel cells [5, 6]. It has been shown that increased gas crossover accelerates membrane degradation [2]. In light of all this, it is vital that one evaluates the gas crossover properties of a proton exchange membrane before they are used in a fuel cell.

Two methods commonly used to measure gas diffusion coefficient and solubility in a polymeric membrane or an ionomer layer are the electrochemical monitoring technique (EMT) [7, 8, 9, 10, 11, 12] and the potential step technique (PST) [13, 14, 15, 16, 17, 18]. The differences in the two methods include: data acquisition time (~15 minutes for the EMT versus ~30 seconds for the PST), electrode size (micro-electrodes are used for the PST compared to electrodes on the order of 1 $cm^2$ for the EMT) and the experimental setup. Further, in PST, the gas is usually dissolved in a liquid electrolyte (i.e. $H_3PO_4$) and diffuses through the ionomer membrane deposited over the working electrode (i.e. rotating disk electrode), while in the EMT, the membrane is directly in contact with the gas-phase. The two methods have been found to agree well with each other qualitatively as shown in **Table 2** for the case of $O_2$ diffusion in Nafion 117.

In spite of above-mentioned important consequences of gas permeability in fuel cell membranes, there is very little relevant data reported in the literature. There are few theoretical methods aimed at understanding gas diffusion through polymeric networks [19] but there are many drawbacks associated with accurately setting up the polymer structure. For example, Nafion membranes are typically operated in the vicinity of their glass transition temperature (Tg) where the rigid frameworks representing their structures do not apply. On the other hand, there isn't a robust experimental setup reported anywhere that can be used to estimate gas crossover properties of a membrane in a functional fuel cell for different fuel cell operating conditions (e.g., pressure. temperature, humidity, etc.). In this study, we report the use of the EMT on an apparatus made from fuel cell assembly to measure the diffusion coefficient and solubility values for $O_2$, CO and $H_2S$ in a Nafion membrane. The setup designed is similar to that of a Devanathan-Stachurski [7] type cell commonly employed to measure hydrogen permeability in metals. The effect of temperature on the diffusion coefficient and solubility of all three gases in Nafion 112 membrane is evaluated and the associated activation energies and enthalpies of mixing are reported. The effect of pressure on the diffusion coefficient and solubility of $O_2$ in Nafion is evaluated using a thicker membrane (Nafion 117). The estimated values are compared to those available in the literature. A similar approach was recently adapted by our group towards characterizing $SO_2$ transport in Nafion 212 and 115 membranes in a functional electrolyzer [20] under open-circuit conditions.

## 2. EXPERIMENTAL

*2.1. Catalyst coated membrane (working electrode) fabrication* – Pt catalyst ink with 75% catalyst and 25% Nafion (dry solids content) was prepared with commercially available 40 wt% Pt on Vulcan XC-72R E-TEK catalyst (PEMEAS Fuel Cell Technologies, Somerset, NJ). Nafion in the form of perfluorosulfonic acid-PTFE copolymer (5% w/w solution, Alfa Aesar, Ward Hill, MA) was used. The catalyst ink was sprayed on to gas diffusion layers (ELAT GDL, 10 $cm^2$ active area, PEMEAS Fuel Cell Technologies, Somerset, NJ) with $N_2$ brush (Paasche Airbrush Company, Chicago, IL), air dried for thirty minutes and then dried under vacuum (381 mm Hg) at 110 °C for 10 minutes to evaporate any remaining solvent. This process was repeated until a catalyst loading of 0.5 mg $Pt/cm^2$ was achieved. One such GDL with the catalyst was then hot pressed onto a pretreated Nafion 112 or 117 membrane (Alfa Aesar, Ward Hill, MA) at 140 °C and 3450 kPa for two minutes to make a catalyst coated membrane. The catalyst-coated membrane obtained using this procedure is analogous to chemical plating procedure reported by Takenaka et al. [21].

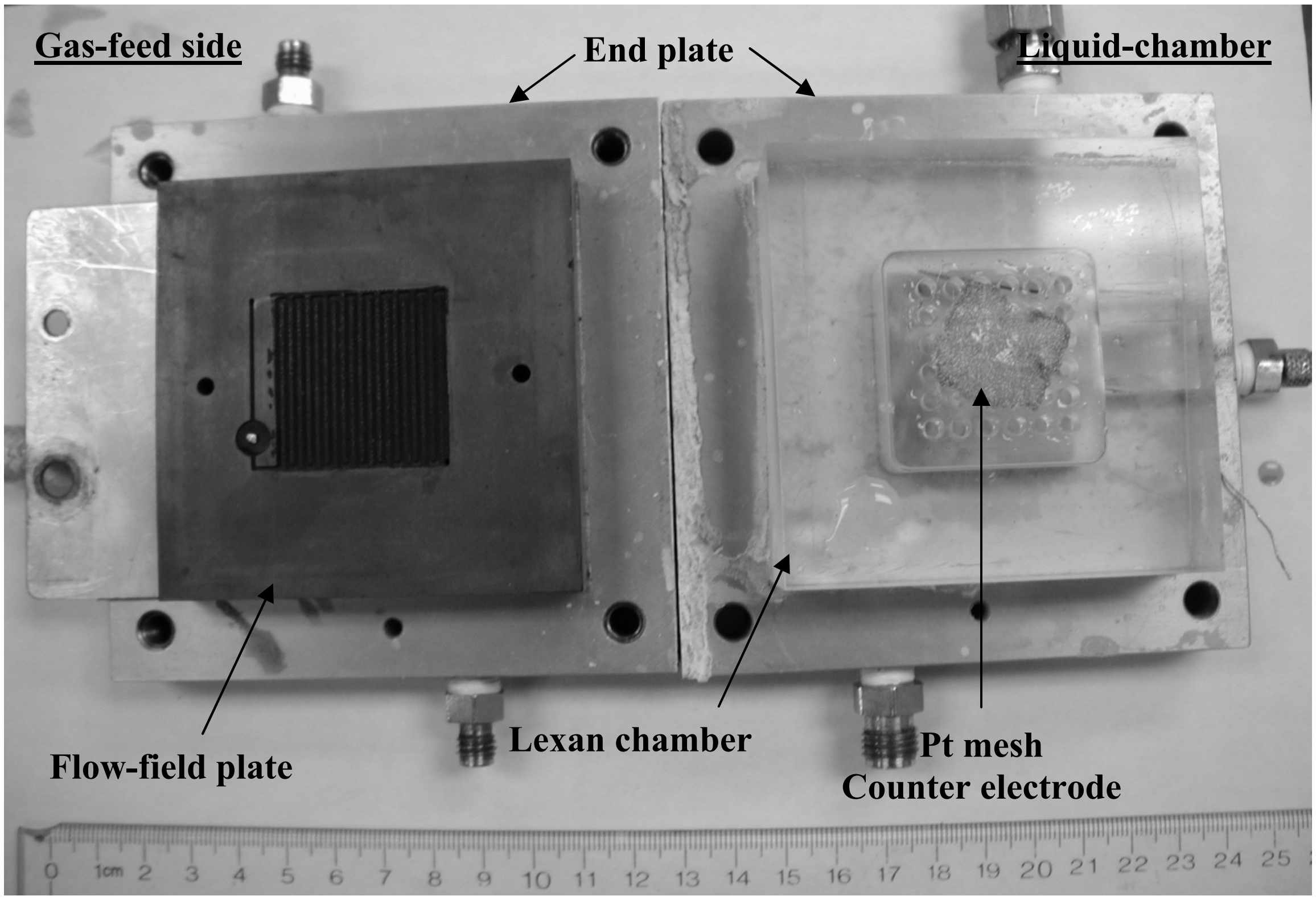

*Figure 1: Photograph of a Devanathan-Stachurski [7] type cell made using a fuel cell assembly for measuring transport and solubility of fuel cell gases in a proton exchange membrane. Shown here is the opened up cell with the gas-feed side on the left and the liquid-chamber on the right. The catalyst coated membrane (working electrode) and the gas diffusion layers are not shown here.*

*2.2. Devanathan-Stachurski cell* – The apparatus used to measure gas diffusion and solubility in a proton exchange membrane was made by modifying an actual fuel cell assembly (Fuel Cell Technologies, Albuquerque, NM) such that one half of the apparatus had gas channels and the other side had a liquid cell. A photograph of the opened up cell (without the electrodes) is shown in Figure 1 and a cross-sectional schematic is shown in Figure 2. This is a direct improvement of a similar glassware apparatus reported earlier by Haug and White [11] such that the effect of gas pressure and temperature can now be studied. The effect of gas humidity on gas transport in the membrane can also be studied using this setup with minimal modifications. The catalyst coated membrane was placed between the gas and liquid cells with appropriate gaskets to prevent leakage. A gas diffusion layer was placed between the gas channels and the catalyst coated membrane to ensure uniform gas distribution. The platinized side of the membrane (working electrode) was facing the liquid cell containing 0.1M $H_2SO_4$ electrolyte, a Pt mesh counter electrode (1 $cm^2$ x 0.2 cm, Alfa Aesar) and saturated KCl-Ag/AgCl reference electrode (~200 mV vs. SHE, Orion Sure-Flow Ag/AgCl Single Junction Half-Cell, Thermo Scientific). The working electrode was held at a constant potential such that the diffusing gas was either oxidized or reduced on its surface.

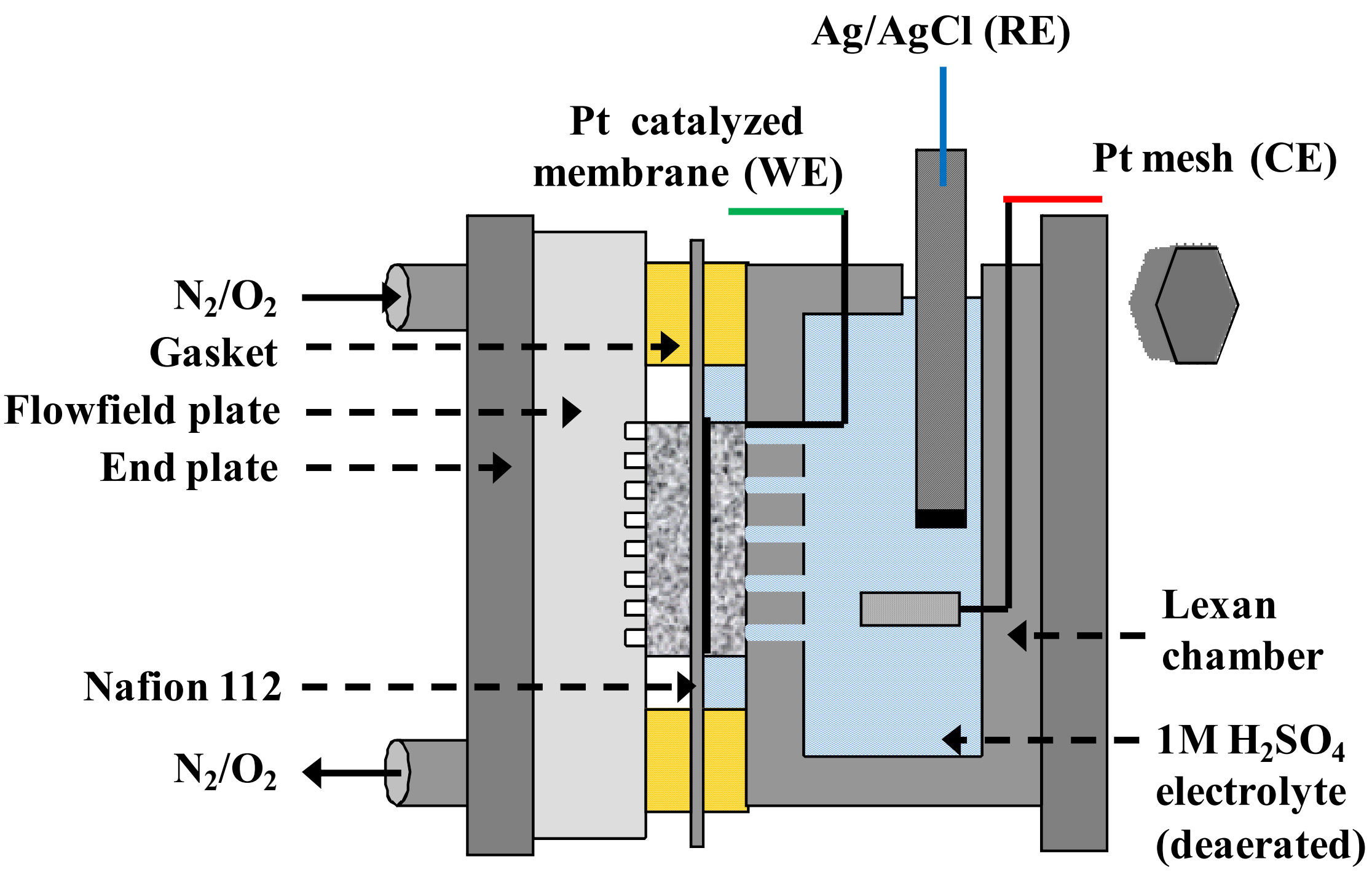

*Figure 2: Schematic (cross-sectional view) of the diffusion cell derived from an actual fuel cell assembly. WE: Working electrode, RE: Reference electrode and CE: Counter electrode.*

An EG&G Princeton Applied Research Potentiostat/Galvanostat (Model 273A, Ametek Inc., Oak Ridge, TN) was used for this purpose. The respective working electrode potentials for $O_2$, CO and $H_2S$ diffusion experiments were 0.1, 0.7 and 0.9 V vs. Ag/AgCl reference electrode. These potentials were chosen such that $O_2$ was reduced at the working electrode while CO and $H_2S$ were oxidized in each of the diffusion experiments. All gases were of ultra high purity grade (Praxair Inc.). Special safety precautions were put in place for handling pure CO and $H_2S$. The gas channels had $N_2$ flowing through them initially while the background current was recorded at the working electrode. Once a steady background current was reached, $N_2$ was replaced by the gas of interest. This was considered the beginning of the diffusion experiment (i.e., t and i are set to zero). An increase in the current on the working electrode was monitored. Data was recorded at a sampling rate of 10 Hz until a new steady-state (limiting current) was reached. This current-time data was used to estimate the diffusion coefficient and the solubility values. The entire experiment was conducted on a fuel cell test station (Fuel Cell Technologies, Albuquerque, NM) made for testing the performances of PEM fuel cells. All the gases fed to the diffusion cell were fully humidified (i.e., the gases were in equilibrium with saturated water vapor at the temperature of the diffusion cell). The temperature of the diffusion cell was controlled by heating elements in the end plates in conjunction with cooling fans on either side. Gas pressure was controlled by a combination of back pressure valves. Between four and ten trials were performed at each set of conditions and an average value of the parameters is reported.

It must be noted that one can measure gas crossover properties of a membrane in an actual functioning PEM fuel cell built with a membrane electrode assembly. Hydrogen crossover measurements routinely conducted as part of membrane durability experiments is one such example [2]. However, the following must be considered: (1) the crossover current measured in such experiments is indicative of gas transport through half the membrane electrode assembly (i.e., microporous layer, catalyst layer and the membrane) and not just the membrane and (2) choosing the correct reference electrode is difficult when measuring the crossover of gases other than hydrogen. To mitigate these concerns and to quickly convert a functioning fuel cell into a diffusion cell, the MEA may be replaced by a catalyst-coated membrane (single-sided) with an addition of an appropriate reference electrode. In addition to this, the cell can be used to study chemical plating inside the membrane by choosing an appropriate electrolyte [e.g., $H_2PtCl_6.(H_2O)_6$] in the liquid cell and a reducing gas (e.g., $H_2$) on the gas channels.

## 3. THEORY

The electrochemical monitoring technique was used to determine the diffusion coefficients and solubilities for gases in membranes. Fick's law and the appropriate boundary conditions presented in equations 1 though 4 were used to define the system:

$$\frac{\partial c(x,t)}{\partial t} = D_g \frac{\partial^2 c(x,t)}{\partial x^2} \qquad 1$$

$$c(x,t) = 0 \quad \text{for} \quad 0 \le x \le L(\lambda) \quad \text{for } t < 0 \qquad 2$$

$$c(x,t) = c_g \quad \text{for} \quad x = 0 \quad \text{for } t \ge 0 \qquad 3$$

$$c(x,t) = 0 \quad \text{for} \quad x = L(\lambda) \quad \text{for } t \ge 0 \qquad 4$$

where $D_g$ is the diffusion coefficient, $c_g$ is the solubility of the diffusing gas, and $L(\lambda)$ is the thickness of the membrane. The steady-state limiting and reaction currents were given by Eq. 5 and 6, respectively,

$$i_\infty = \frac{n_{e,g}}{s_g}\frac{FAD_g c_g}{L(\lambda)} \quad 5$$

$$i(t) = \frac{n_{e,g}}{s_g} FAD_g \left.\frac{\partial c(x,t)}{\partial x}\right|_{x=L(\lambda)} \quad 6$$

where $n_{e,g}$ is the number of electrons in the electrochemical reaction of interest corresponding to reduction or oxidation of gas g, $s_g$ is the stoichiometric coefficient of the diffusing gas, $L(\lambda)$ is the membrane thickness, A is the cross sectional area of the working electrode, and F is Faraday's constant. The membrane of interest in this case is Nafion and since Nafion swells upon water uptake, the diffusion length, $L(\lambda)$, is a function of water content [22] and is given as,

$$L(\lambda) = L^0\left(1 + 0.36\frac{\hat{\lambda}\bar{V}_0}{\bar{V}_m}\right) \quad 7$$

where $\lambda$ is the ratio of moles of water per mole of sulfonic acid sites in the membrane (i.e., water content), $L^0$ is the dry membrane thickness, $\hat{\lambda}$ is the average water content in the membrane, $\bar{V}_0$ is the initial volume of the membrane and $\bar{V}_m$ is the partial molar volume of the membrane. The catalyst layer is assumed to be infinitesimally thin and swelling of ionomer in the catalyst is ignored. The partial molar volume is obtained from the ratio of the molecular weight of the membrane and its density ($\bar{V}_m = M_m/\rho_m$). Myers and Newman have treated swelling in this manner to ensure conservation of membrane mass [23]. The Devanathan-Stachurski type setup used to measure gas diffusion in this work (see experimental section) has the membrane in equilibrium with 0.1M $H_2SO_4$ on one side and fully humidified gas on the other side. Therefore, $\lambda$ is 22 (assuming water activity to be same in 0.1M $H_2SO_4$ as in pure liquid water) on the membrane-liquid interface and 14 on the membrane-gas interface [24] and if one assumes a linear profile for $\lambda$ inside the membrane, the value for $\hat{\lambda}$ results in 18. This assumption is in accord with the experimental observation by Morris and Sun [25]. Temperature effect on membrane water uptake is ignored. The new diffusion lengths, therefore, are 50.83 μm and 177.86 μm for Nafion 112 and 117 membranes respectively.

The electrochemical reactions of interest for $O_2$ [26], CO [27] and $H_2S$ [28], respectively are,

$$\tfrac{1}{2}O_2 + 2H^+ + 2e^- \rightleftharpoons H_2O, \qquad E^0 = 1.229 \text{ V vs. SHE} \quad 8$$

$$CO + H_2O \rightleftharpoons CO_2 + 2H^+ + 2e^-, \quad E^0 = 0.7 \text{ vs. SHE} \quad 9$$

$$H_2S + 2H_2O \rightleftharpoons SO_2 + 6H^+ + 6e^-, \quad E^0 = 0.9 \text{ vs. SHE} \quad 10$$

Fan [29] solved the above system of equations using the Laplace transform techniques resulting in Eq. **11**,

$$i(t) = \frac{n_{e,g}FAD_g c_g}{s_g L(\lambda)}\left(\frac{2}{\sqrt{\pi\tau}}\sum_{j=0}^{\infty}\exp\left[-\frac{(2j+1)^2}{4\tau}\right]\right) \quad 11$$

Where,

$$\tau = \frac{tD_g}{L(\lambda)^2} \quad 12$$

Equation **11** can be used with data for i(t) to obtain $D_g$ and $c_g$ as discussed below.

The method of least squares [30] was used to fit the data from each trial to Eq. **11** and solve for the diffusion coefficient and solubility simultaneously. To determine the accuracy of values obtained for D and $c_g$, confidence intervals were obtained by using the method described by Kimble and White [31] shown in Eq. **13**,

$$P_k = \hat{P}_k \pm t_\gamma s_{\hat{P}_k}\sqrt{C_{kk}} \quad 13$$

where $\hat{P}_k$ is the estimate of parameter $P_k$ found through the least squares method, $s_{\hat{P}_k}$ is the standard deviation for the data set, and $t_\gamma$ is the value of the t-distribution (also known as the student distribution) [10,32,33] with a confidence, $\gamma$. Equation **14** is solved for $t_\gamma$ to obtain the t-distribution,

$$\int_{t_\gamma}^{\infty}\frac{\Gamma\left[(f-1)/2\right]}{\sqrt{\pi f}\;\Gamma(f/2)}\left(1+\frac{x^2}{f}\right)^{-\frac{f+1}{2}}dx = \alpha \quad 14$$

$$\alpha = (1-\gamma)/2 \quad 15$$

where f is the degrees of freedom and is equal to (n – m), where n is the number of data points and m is the number of parameters (two in this case, $D_g$ and $c_g$).

A value for $C_{kk}$ in Eq. **13** can be obtained from the approximate Hessian Matrix [31],

$$N=\begin{bmatrix} 2\sum_{j=1}^{n}\frac{\partial i(j)}{\partial P_{D_g}}\frac{\partial i(j)}{\partial P_{D_g}} & 2\sum_{j=1}^{n}\frac{\partial i(j)}{\partial P_{D_g}}\frac{\partial i(j)}{\partial P_{c_g}} \\ 2\sum_{j=1}^{n}\frac{\partial i(j)}{\partial P_{c_g}}\frac{\partial i(j)}{\partial P_{D_g}} & 2\sum_{j=1}^{n}\frac{\partial i(j)}{\partial P_{c_g}}\frac{\partial i(j)}{\partial P_{c_g}} \end{bmatrix} \quad 16$$

where i(j) is the current, i, recorded at each data point, j. Equation **16** is then inverted and the diagonal elements of that matrix, **N**[1,1] and **N**[2,2], are taken as $C_{kk}$ ($C_{D_gD_g}$ for diffusivity and $C_{c_gc_g}$ for solubility).

Since the membrane is in equilibrium with 0.1 M $H_2SO_4$ on one side and fully humidified gas on the other, the resulting water flux and the associated transport of dissolved gas from the liquid side to the gas side (counter to the direction of diffusion) need to be quantified. Nguyen and White [34] report the Fickian diffusion coefficient [35] for water in Nafion as,

$$D_{W,F} = (1.76 \times 10^{-5} + 1.94 \times 10^{-4}\lambda) \exp\left[\frac{-2436}{T}\right] \qquad 17$$

The permeability of the gases studied (namely $O_2$, CO and $H_2S$) due to water transport in the direction counter to the direction of gas diffusion can then be given as,

$$P_{g,W} = X_{g,W} c_W D_{W,F} \qquad 18$$

where $X_{g,W}$ is the mole fraction solubility of gas g (g = $O_2$, CO or $H_2S$) in pure liquid water, $c_W$ is the solubility of water in Nafion membrane. The mole fraction solubility of the $O_2$ [36], CO [37] and $H_2S$ [38] are respectively calculated using the following correlations,

$$\ln\left(X_{O_2,W}\right) = A_1 + \frac{B_1}{T^*} + C_1 \ln T^*,\ T^* = \frac{T}{100} \qquad 19$$

$$\ln\left(X_{CO,W}\right) = A_2 + A_3\frac{100}{T} + A_4 \ln\left(\frac{T}{100}\right) + A_5\frac{T}{100} + S\left[B_2 + B_3\frac{T}{100} + B_4\left(\frac{1}{100}\right)^2\right] \qquad 20$$

$$\ln\left(X_{H_2S,W}\right) = A_6 + A_7 T + A_8 T^2 + \frac{B_5}{T} + C_2 \ln T \qquad 21$$

$A_n$, $B_n$ and $C_n$ are coefficients and their values are given in Table 1. The permeability values estimated for the gases from the gas side to the liquid side is corrected for the above. Activation energies for diffusion and mixture enthalpies for the gases were estimated by fitting the estimated diffusion coefficient and solubility values, respectively, to the following Arrhenius relationships,

$$D_g = D_g^0 \exp\left[\frac{E_{a,g}}{RT}\right] \qquad 22$$

$$c_g = c_g^0 \exp\left[\frac{H_g}{RT}\right] \qquad 23$$

where $E_{a,g}$ and $H_g$ are respectively the activation energy for gas diffusion and mixture enthalpy for gas g in a Nafion membrane. Maple 10 (Maplesoft, Waterloo, Ontario) was used to run the parameter estimation routines. Values for parameters used in the data analysis are listed in Table 1.

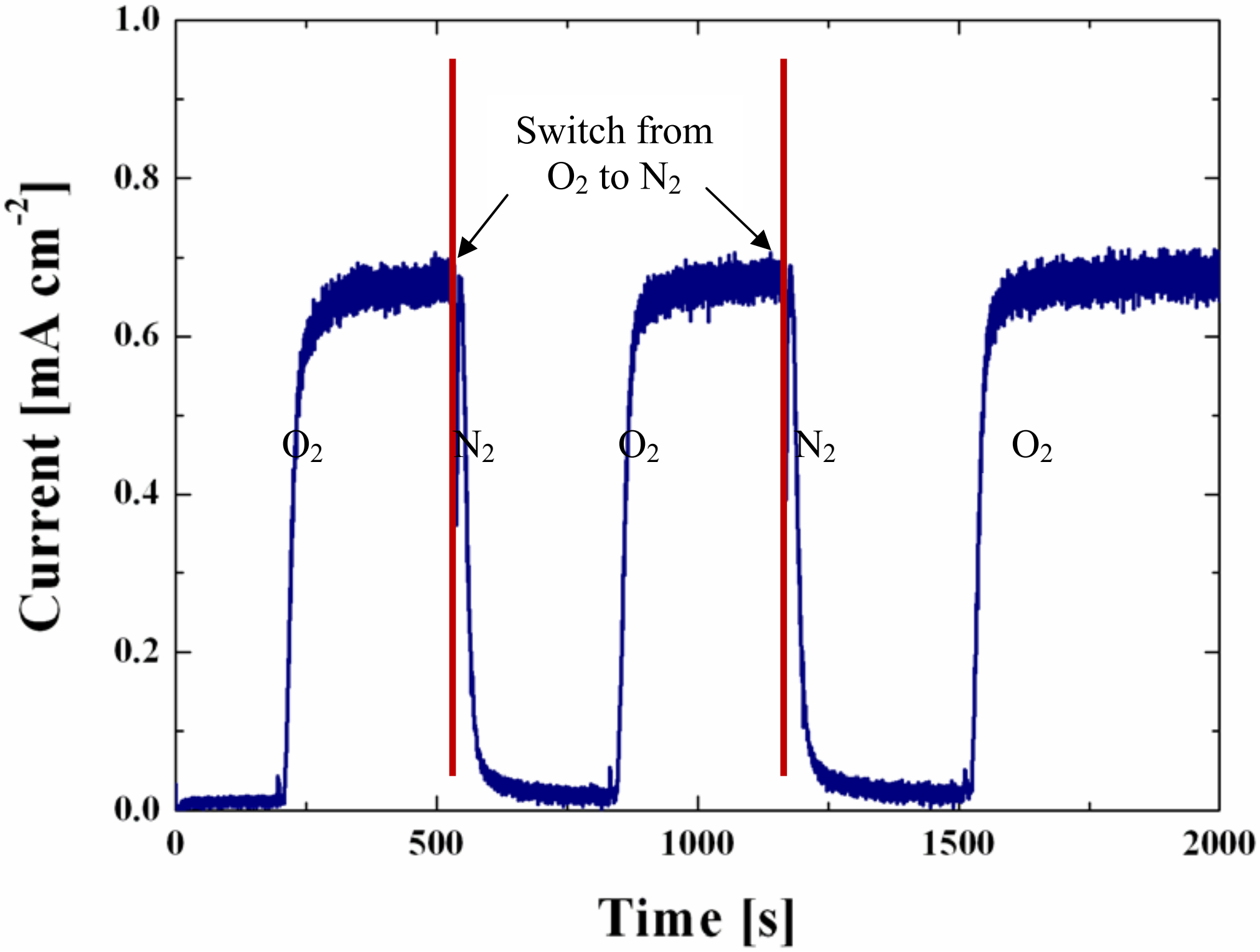

*Figure 3: Background corrected chronoamperometric response at the working electrode for switches between $N_2$ and $O_2$ on the gas side at 35 °C. The rise in current followed by a stable value is indicative of $O_2$ diffusion in the membrane (Nafion 112) and reduction to $H_2O/H_2O_2$. The working electrode was held at 0.1 V vs. Ag/AgCl reference electrode. This current response from the working electrode was used in conjunction with equation 11 to extract $O_2$ diffusion coefficient and solubility in Nafion 112.*

## 4. RESULTS AND DISCUSSION

*4.1. Oxygen ($O_2$)* – ***Figure 3*** shows the background corrected current-time data obtained at the working electrode due to $O_2$ diffusion through a Nafion 112 membrane followed by its reduction according to equation **8**. Three switches between $N_2$ and $O_2$ are shown. The area under the curve marked $O_2$ or $N_2$ in the figure represents the corresponding gas in the bulk on the gas-channels side of the diffusion cell. It can be seen that the rise time and the steady-state current ($i_\infty$) are uniform between the switches. **Table 2** lists values for diffusion coefficient and solubility values for $O_2$ in the literature for various temperatures and pressures. The values reported by Haug and White [11] were used as the initial guess for the parameter estimation routine based on $O_2$ diffusion measurements in this study.

*4.1.1. Effect of temperature* – The diffusion measurements for $O_2$ were conducted as a function of temperature for Nafion 112 and Nafion 117 membranes and the current response at the working electrode is plotted in ***Figure 4***.

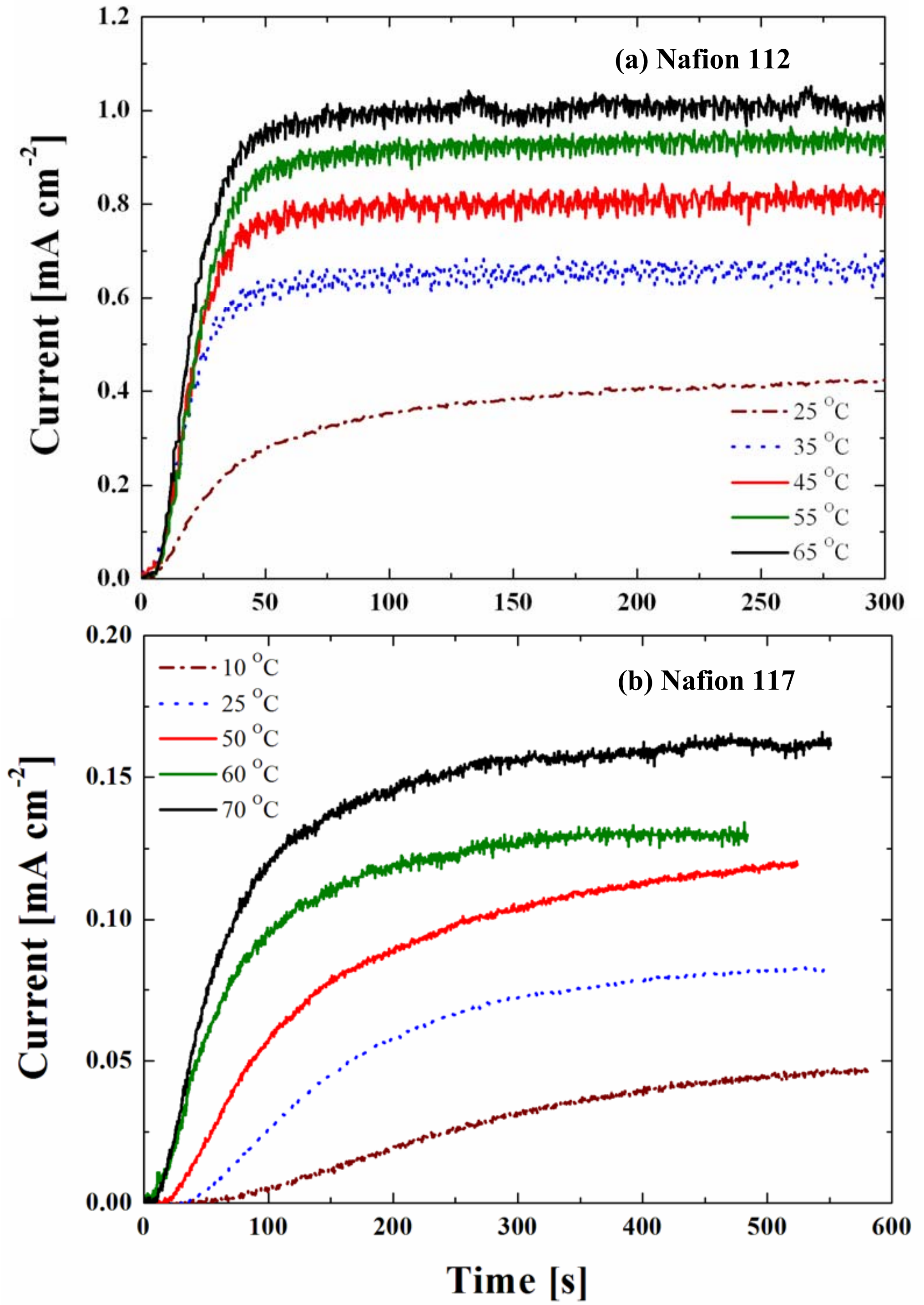

*Figure 4: Current response at the working electrode due to oxygen diffusion through (a) Nafion 112 and (b) Nafion 117 and subsequent reduction at the working electrode at different temperatures. The working electrode was held at 0.1 V vs. Ag/AgCl reference electrode (~0.3 V vs. SHE).*

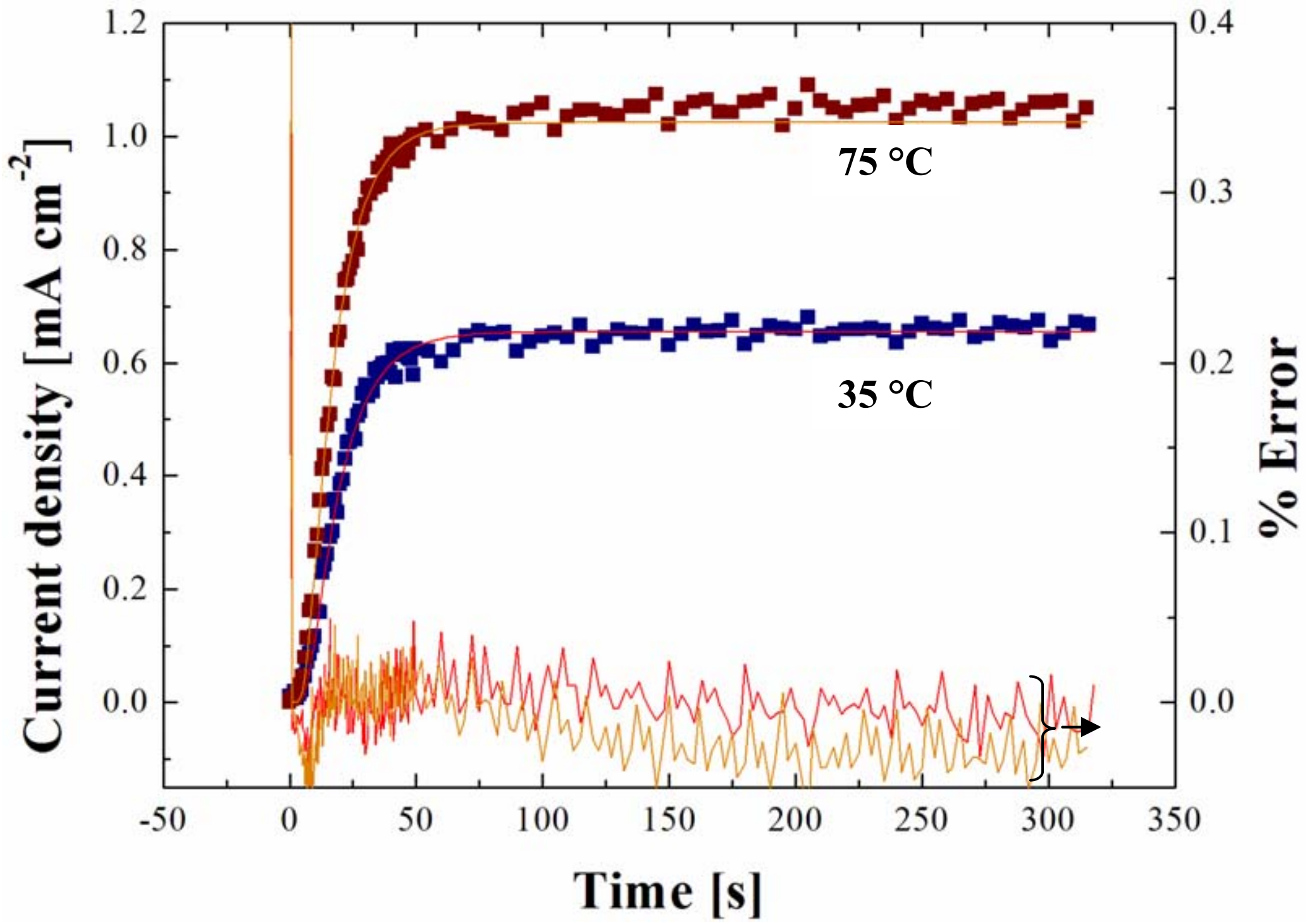

*Figure 5: Comparison of the experimental data (symbols) to equation 11 (lines) fitted with the diffusion coefficient and solubility of $O_2$ in Nafion 112 (for 25 and 65 °C) determined from the method of least squares. The respective error between data and the fit is shown on the secondary axis.*

Because of the longer diffusion length, the evolution of current for the case of Nafion 117 (***Figure 4***b) is slower and distinct with temperature compared to that of Nafion 112 (***Figure 4***a). For reasons unknown, the noise to signal ratio is considerably larger for the Nafion 112 membrane. As can be seen, the permeability of $O_2$ (proportional to the steady-state current) increases with temperature for both membranes. The model fits the data well over the entire temperature range for both membranes. Comparison between data and fit from equation **11** is shown for two temperatures for Nafion 112 in ***Figure 5***. The corresponding error values indicate that the disparity between the model and the data are high during the transience while it is mostly noise during steady-state. The resulting diffusion coefficient and solubility values for various temperatures for Nafion 112 and 117 membranes are plotted in ***Figure 6*** and ***Figure 7***, respectively. The corresponding confidence intervals and the permeability values for Nafion 112 and Nafion 117 are listed in Table 3a and Table 3b, respectively. Oxygen permeability values estimated in this work by the EMT technique is similar to those estimated by Broka and Ekdunge [39] using gas chromatography; by Chiou and Paul [40] as well as by Sakai et al. [41, 42] using volumetric methods with high pressure permeation cells. The exponential fits to data obtained in this work as shown in ***Figure 6*** correspond to the following,

$$D_{O_2,\,\text{Nafion 112}} = 17.45 \times 10^{-6} \exp\left[\frac{-1514}{T}\right] \qquad 24$$

$$D_{O_2,\,\text{Nafion 117}} = 24.82 \times 10^{-5} \exp\left[\frac{-1949}{T}\right] \qquad 25$$

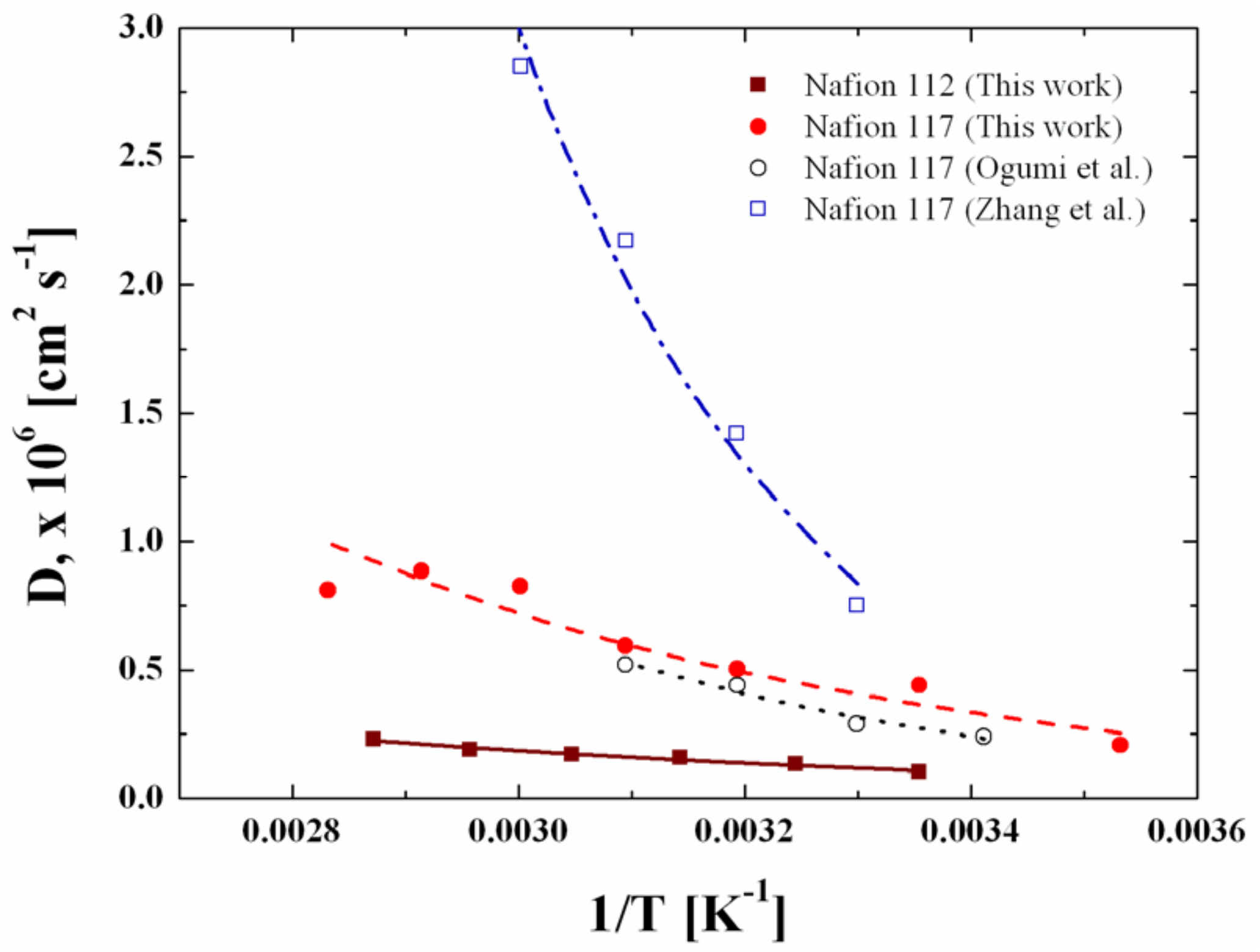

*Figure 6: Oxygen diffusion coefficient in Nafion 112 and 117 estimated using the electrochemical monitoring technique is compared to those reported by Ogumi et al. [9] and Zhang et al. [43]. The symbols correspond to data and the lines correspond to exponential fits. For data obtained in this work the fits correspond to equations 24 and 25. The estimated activation energies for $O_2$ diffusion through Nafion 112 and Nafion 117 are ~12.58 kJ mol$^{-1}$ and ~16.2 kJ mol$^{-1}$ respectively.*

The corresponding activation energies for $O_2$ diffusion in Nafion 112 and 117 are 12.58 kJ mol$^{-1}$ and 16.2 kJ mol$^{-1}$ respectively. The exponential fits to data obtained in this work as shown in ***Figure 7*** correspond to the following,

$$c_{O_2,\,\text{Nafion 112}} = 10.29 \times 10^{-4} \exp\left[\frac{-707.5}{T}\right] \qquad 26$$

$$c_{O_2,\,\text{Nafion 117}} = 2.41 \times 10^{-6} \exp\left[\frac{-605}{T}\right] \qquad 27$$

The corresponding enthalpies of mixture for $O_2$ in Nafion and Nafion 117 are 5.88 kJ mol$^{-1}$ and 5.03 kJ mol$^{-1}$ respectively.

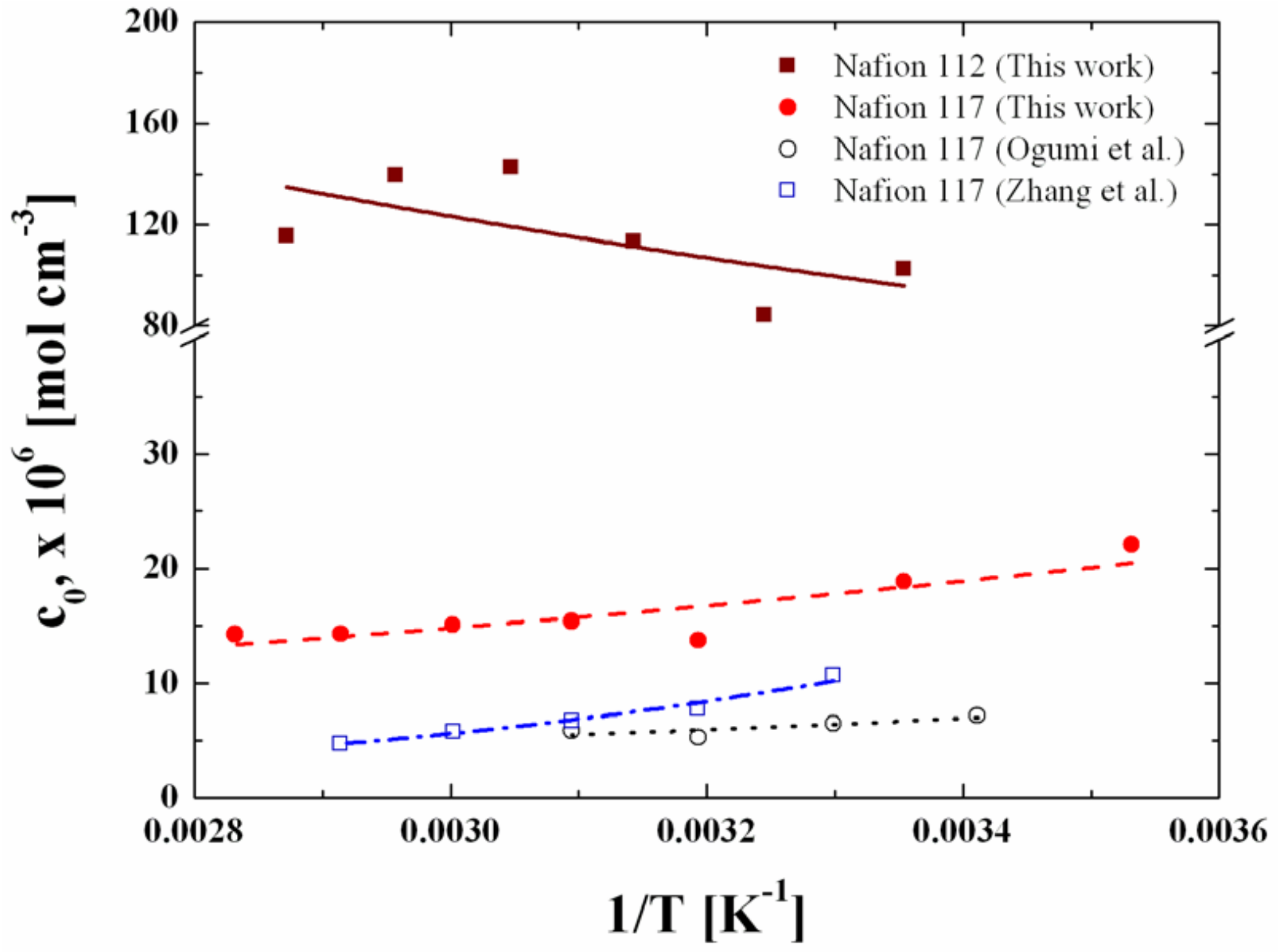

*Figure 7: Oxygen solubility in Nafion 112 and 117 estimated using the electrochemical monitoring technique is compared to those reported by Ogumi et al. [9] and Zhang et al. [43]. The symbols and lines correspond to data and exponential fits, respectively. For data obtained in this work, the fits correspond to equations 26 and 27. The estimated mixture of enthalpies for Nafion 112 and 117 are 5.88 kJ mol$^{-1}$ and 5.03 kJ mol$^{-1}$ respectively.*

*4.1.2. Effect of pressure* – The diffusion cell was used to evaluate the effect of $O_2$ pressure on its permeability in Nafion 117. A thicker membrane was chosen to withstand the pressure differential since the liquid chamber was not pressurized. The electrochemical response at the working electrode is shown for different $O_2$ pressures from 1 atm to 3 atm in ***Figure 8*** and the corresponding diffusion coefficient, solubility and permeability values along with the confidence intervals are given in Table 3c. The model fit the data very well over the entire pressure range. The diffusion coefficient of $O_2$ in Nafion was invariant with $O_2$ pressures between 1 atm and 3 atm. However, the solubility of $O_2$ in Nafion increased linearly with pressure and followed Henry's law for dilute gases (since the amount of oxygen dissolved in Nafion is very low),

$$K_{O_2} = \frac{P_{O_2}}{x_{O_2}} \qquad 28$$

where $P_{O_2}$ is the partial pressure of $O_2$ and x is the mole fraction of $O_2$ in Nafion. The resulting Henry's law constant is 3.504 x 10$^3$ atm. Qualitatively, the trend of increasing $O_2$ solubility and an unchanging $O_2$ diffusion coefficient with pressure agree well with the results reported by Beattie et al. [15] for Nafion 117.

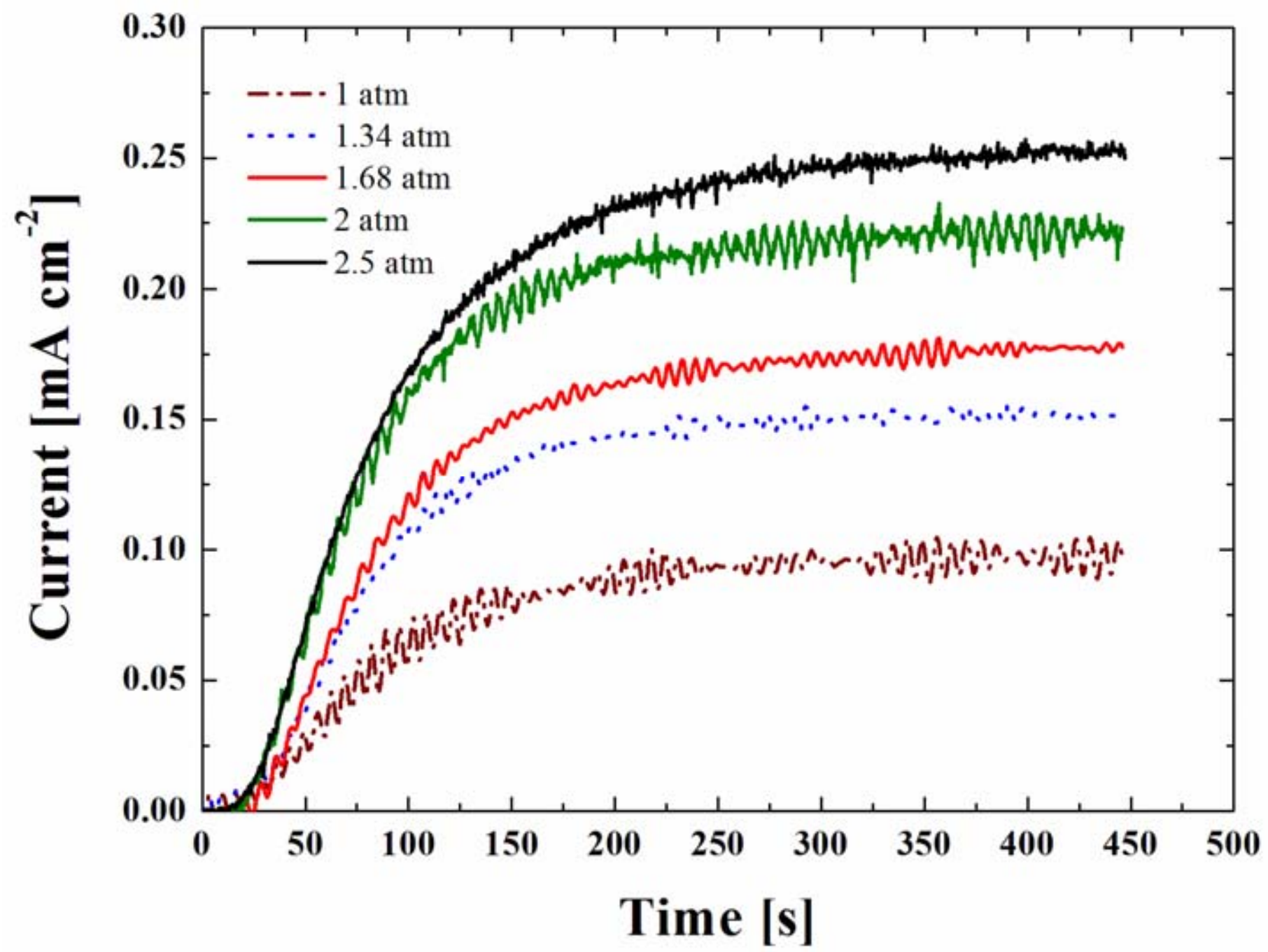

*Figure 8: Current response due to oxygen diffusion through Nafion 117 membrane and reduction at the working electrode for different oxygen pressures from 1 to 3 atm and 25 °C. The working electrode was held at 0.1 V vs. Ag/AgCl reference electrode (~0.3 V vs. SHE).*

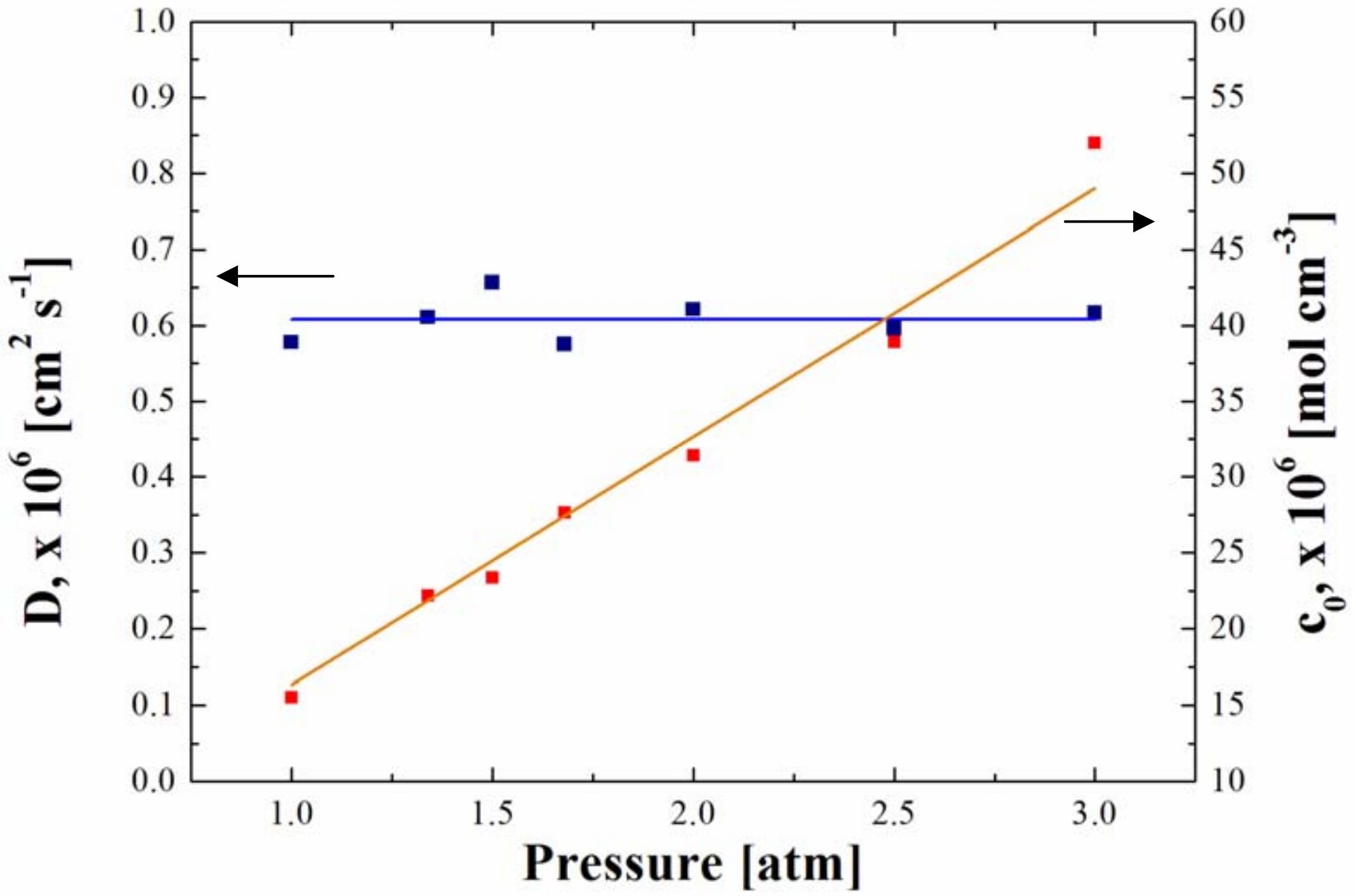

*Figure 9: Diffusion coefficient and solubility of oxygen in Nafion 117 for different oxygen pressures between 1 and 3 atm at 25 °C. The symbols and lines respectively correspond to data and linear fits.*

*4.2. Carbon monoxide (CO)* – Though there is plenty of data reported in the literature on $O_2$ diffusion in Nafion, there aren't any analogous measurements on CO diffusion. This is not because of lack of need for such parameters. This is because, one can always assume a CO diffusion coefficient value equal to that of $O_2$ since the size of a CO molecule is only 14% smaller than that of an $O_2$ molecule. These values are routinely used by the PEM fuel cell modeling community to simulate the behavior of unit cells or fuel cell stacks operating on reformate feed that have as much as 500 ppm CO. We therefore attempt to experimentally obtain these parameters using the EMT.

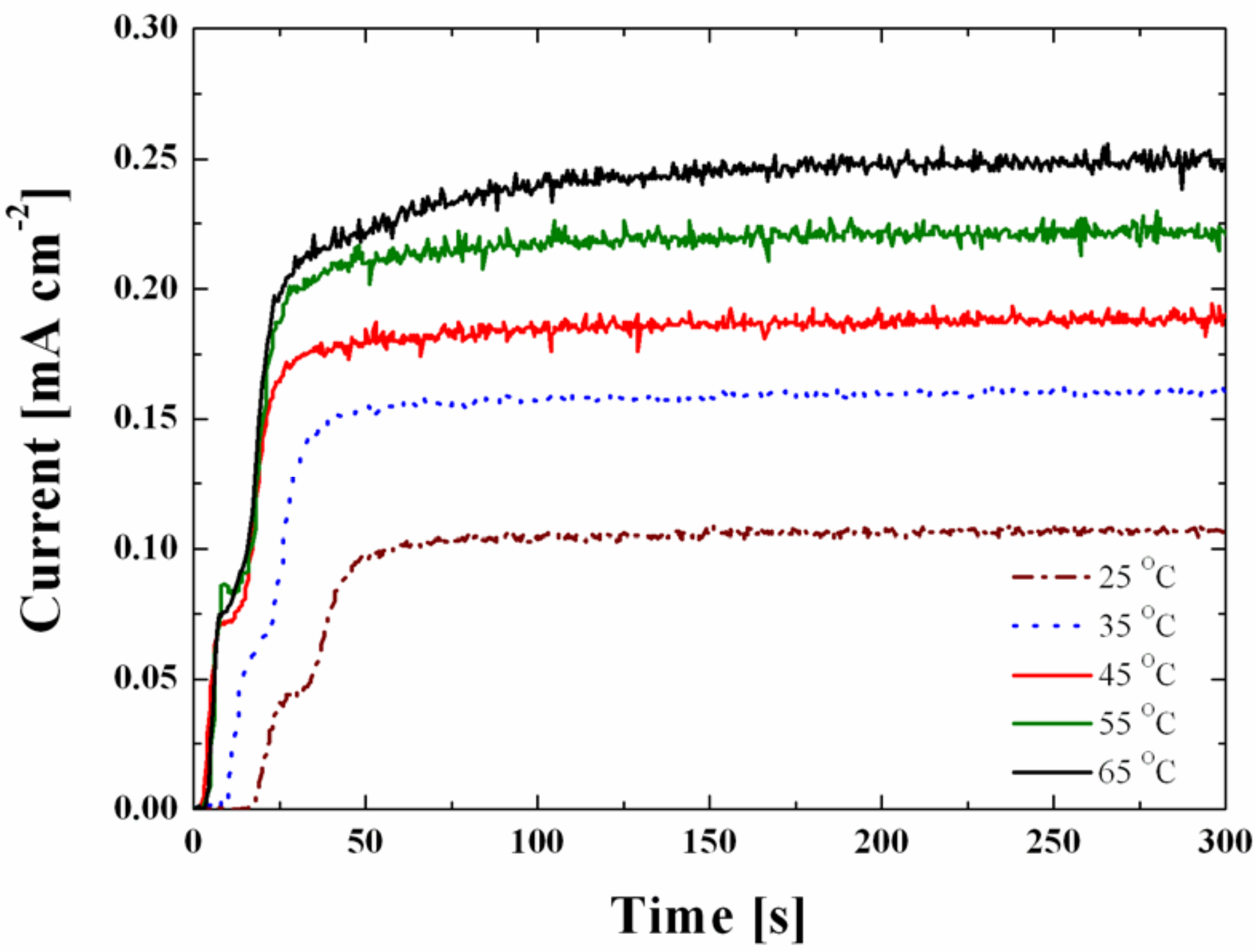

*Figure 10: Current response due to CO diffusion through Nafion 112 and subsequent oxidation to $CO_2$ at the working electrode at different temperatures. The working electrode was held at 0.9 V vs. Ag/AgCl (~1.1 V vs. SHE).*

*4.2.1. Effect of temperature* – ***Figure 10*** shows the current response of the working electrode as a function of temperature between 25 and 65 °C. Unlike $O_2$, the CO current transience has two distinct features – the time constants are remarkably different even for Nafion 112 and the current transience shows a step like behavior. Since no such behavior was seen for the case of $O_2$ diffusion, dual diffusion pathways can be ruled out. This could however be indicative of the way CO electro-oxidation occurs at the working electrode. The model fits the data well except during the transience where the step-like behavior is seen. The fit is shown in ***Figure 11*** where the large error values in the neighborhood of the step indicative of this can be seen. The estimated diffusion coefficient and solubility values as a function of temperature are plotted in ***Figure 12***.

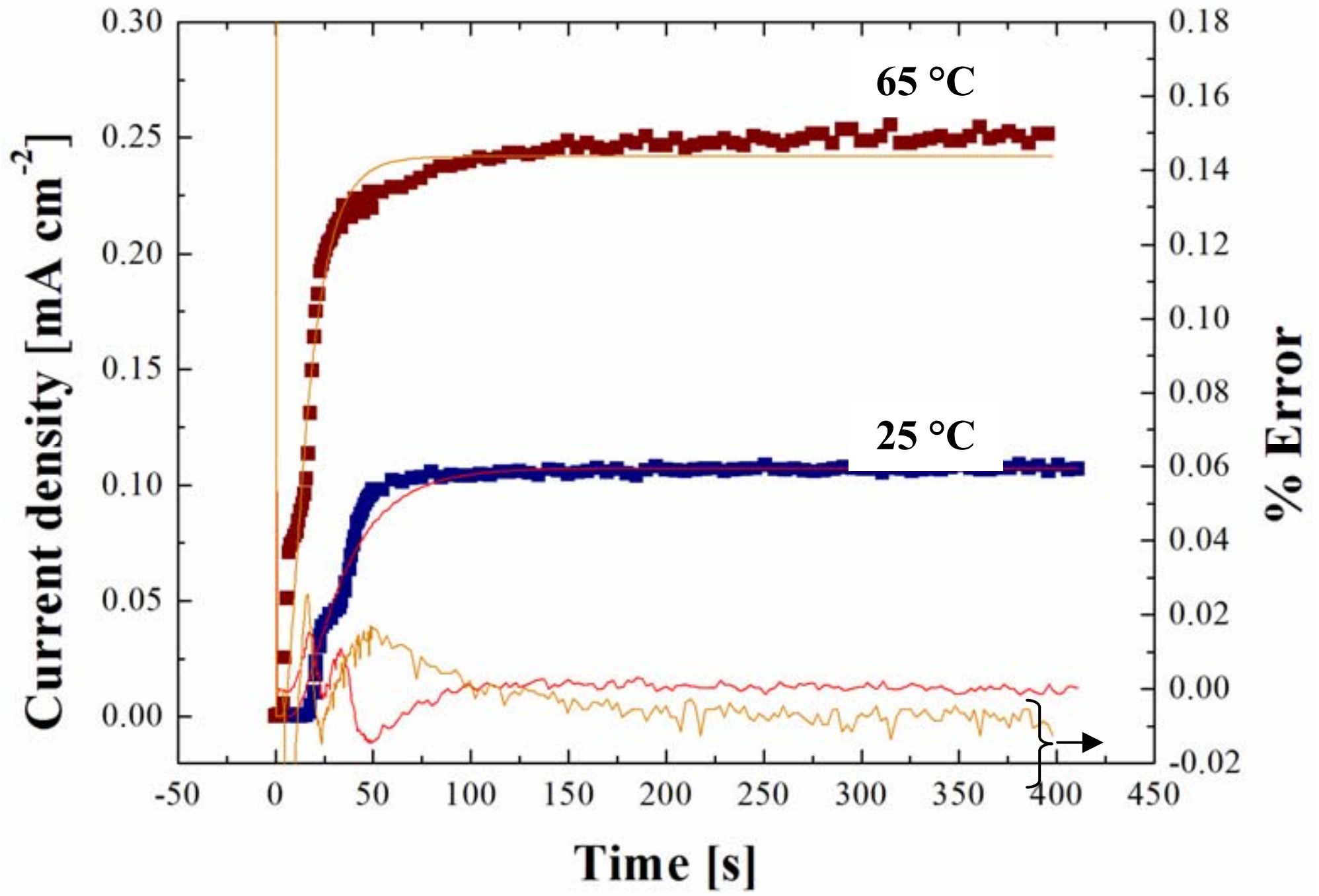

*Figure 11: Comparison of the experimental data (symbols) to equation 11 (lines) fitted with the diffusion coefficient and solubility of CO in Nafion 112 (for 25 and 65 °C) determined from the method of least squares. The respective error between data and the fit is shown on the secondary axis.*

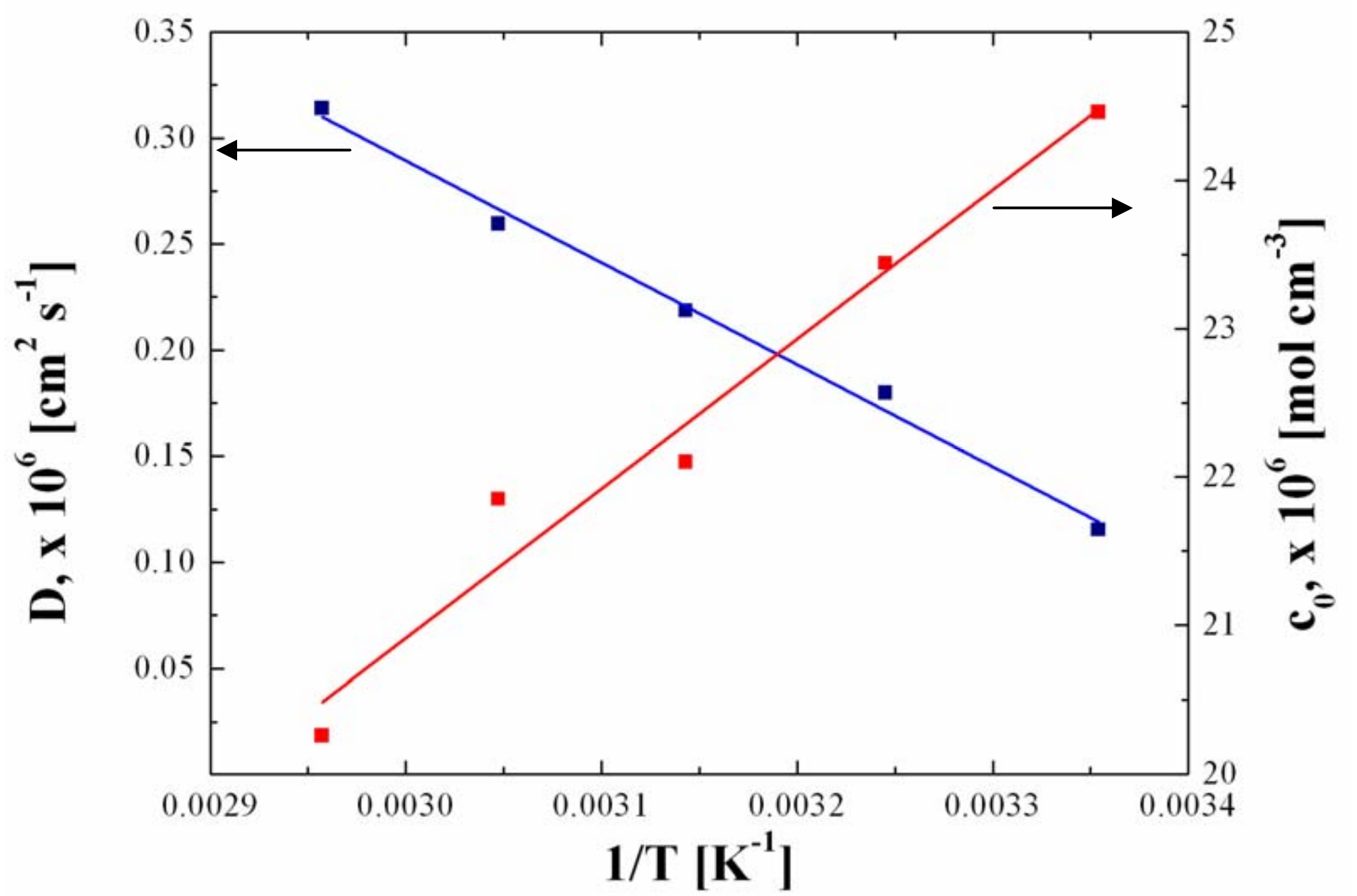

*Figure 12: Carbon monoxide diffusion coefficient and solubility in Nafion 112 estimated from data obtained using the electrochemical monitoring technique. The lines are exponential fits according to equations 29 and 30. The estimated activation energy for CO diffusion through Nafion 112 is ~20kJ mol$^{-1}$ and the mixture of enthalpy is ~3.74 kJ mol$^{-1}$.*

The resulting Arrhenius equations for diffusion coefficient and solubility of CO in Nafion 112, respectively are,

$$D_{CO,\ Nafion\ 112} = 4.02 \times 10^{-4} \exp\left[\frac{-2406}{T}\right] \quad 29$$

$$c_{CO,\ Nafion\ 112} = 5.43 \times 10^{-6} \exp\left[\frac{449.8}{T}\right] \quad 30$$

The corresponding activation energy for CO diffusion through Nafion is ~20 kJ $mol^{-1}$ and the mixture of enthalpy is ~3.74 kJ $mol^{-1}$.

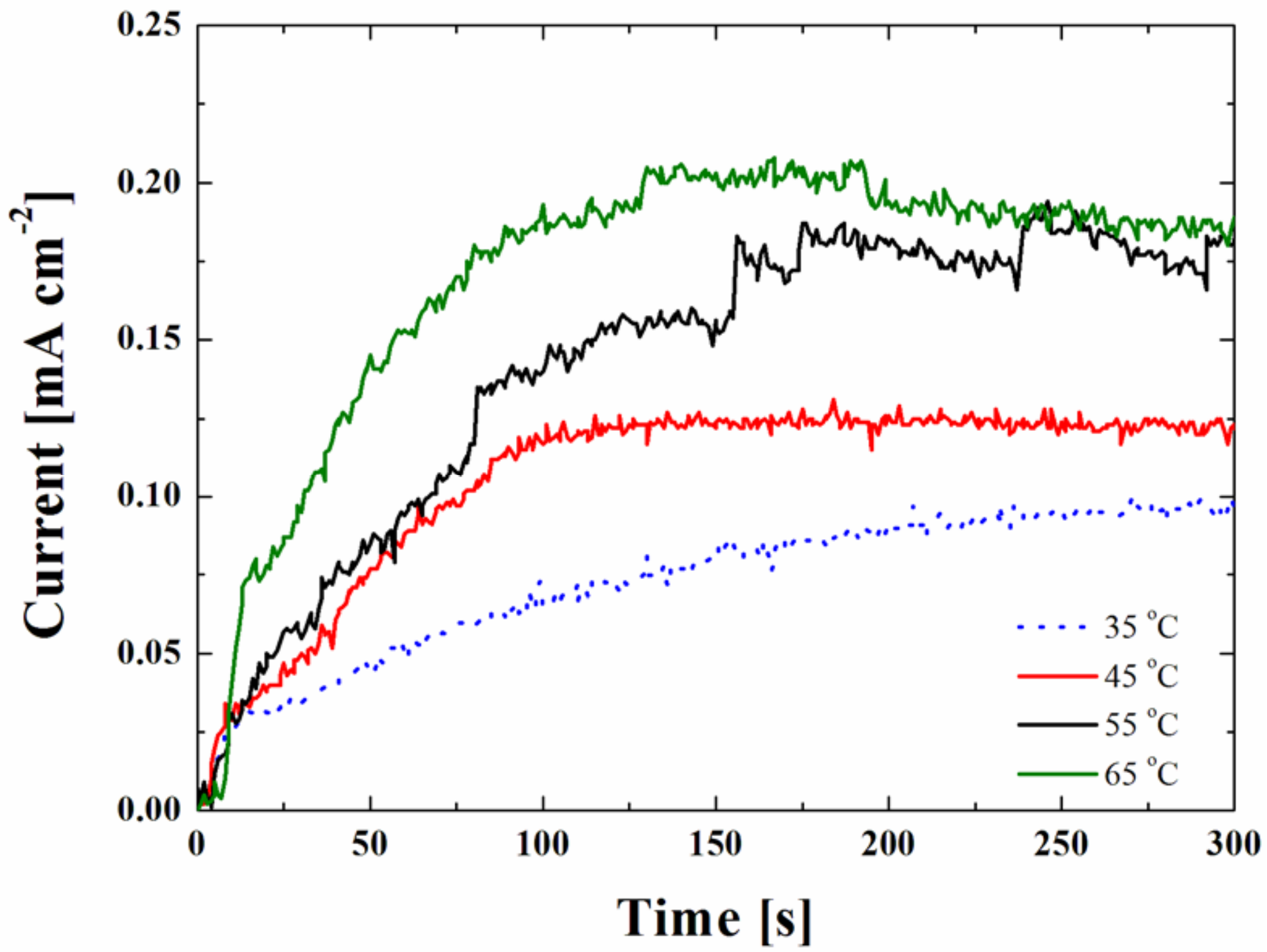

*Figure 13: Current response due to $H_2S$ diffusion through Nafion 112 and subsequent oxidation at the working electrode at different temperatures. The working electrode was held at 0.9 V vs. Ag/AgCl (~1.1 V vs. SHE).*

*4.3. Hydrogen sulfide ($H_2S$)* – $H_2S$ ionization ( $H_2S \underset{}{\overset{k_s}{\rightleftharpoons}} H^+ + HS^-$ ) in water was taken into account and was assumed to be in equilibrium with the dissociated species. The ionization constant for the first dissociation of $H_2S$ is 3.9 x $10^{-8}$ at 0 °C and 3.0 x 10-7 at 100 °C and is assumed to be linear in this range [44]. The second ionization constant was reported in a review by Myers [45] to be ~$10^{-19}$. Due to the exceedingly small value for the second ionization constant, we assume that all of $H_2S$ in the aqueous phase and in Nafion are in molecular form. The overall diffusion of all the species is measured in this study and therefore $H_2S$ diffusion refers to the diffusion of $H_2S$ and its dissociated species.

*4.3.1. Effect of temperature* – ***Figure 13*** shows the current response at the working electrode due to $H_2S$ diffusion through Nafion 112 and subsequent oxidation at the working electrode. The data was fit to equation **11** and is shown in ***Figure 14***. It can be seen from the large error values that the fit is not good in the transient region. Since $H_2S$ dissolves in water, the combined diffusion of the dissociated species and their oxidation is thought to physically take place as one fixed group and therefore treated mathematically with one set of equations. This can lead to significant errors in estimation of the diffusion coefficient and solubility values. It is assumed that the diffusion coefficient of $H_2S$ in Nafion 112 is same as that of $HS^-$, which may not be true since the size of latter is 3% smaller than the former. The estimated diffusion coefficient and solubility values as a function of temperature are plotted in ***Figure 15***. The resulting Arrhenius equations for diffusion coefficient and solubility of $H_2S$ in Nafion 112, respectively are,

$$D_{H_2S,\ \text{Nafion 112}} = 2.87 \times 10^{-6} \exp\left[\frac{-1065}{T}\right] \qquad 31$$

$$c_{H_2S,\ \text{Nafion 112}} = 5.86 \times 10^{-4} \exp\left[\frac{-915}{T}\right] \qquad 32$$

The corresponding activation energy for $H_2S$ diffusion through Nafion is ~8.85 kJ mol$^{-1}$ and the mixture of enthalpy is ~7.61 kJ mol$^{-1}$. Unlike $O_2$ and CO, solubility of $H_2S$ in Nafion increases with temperature.

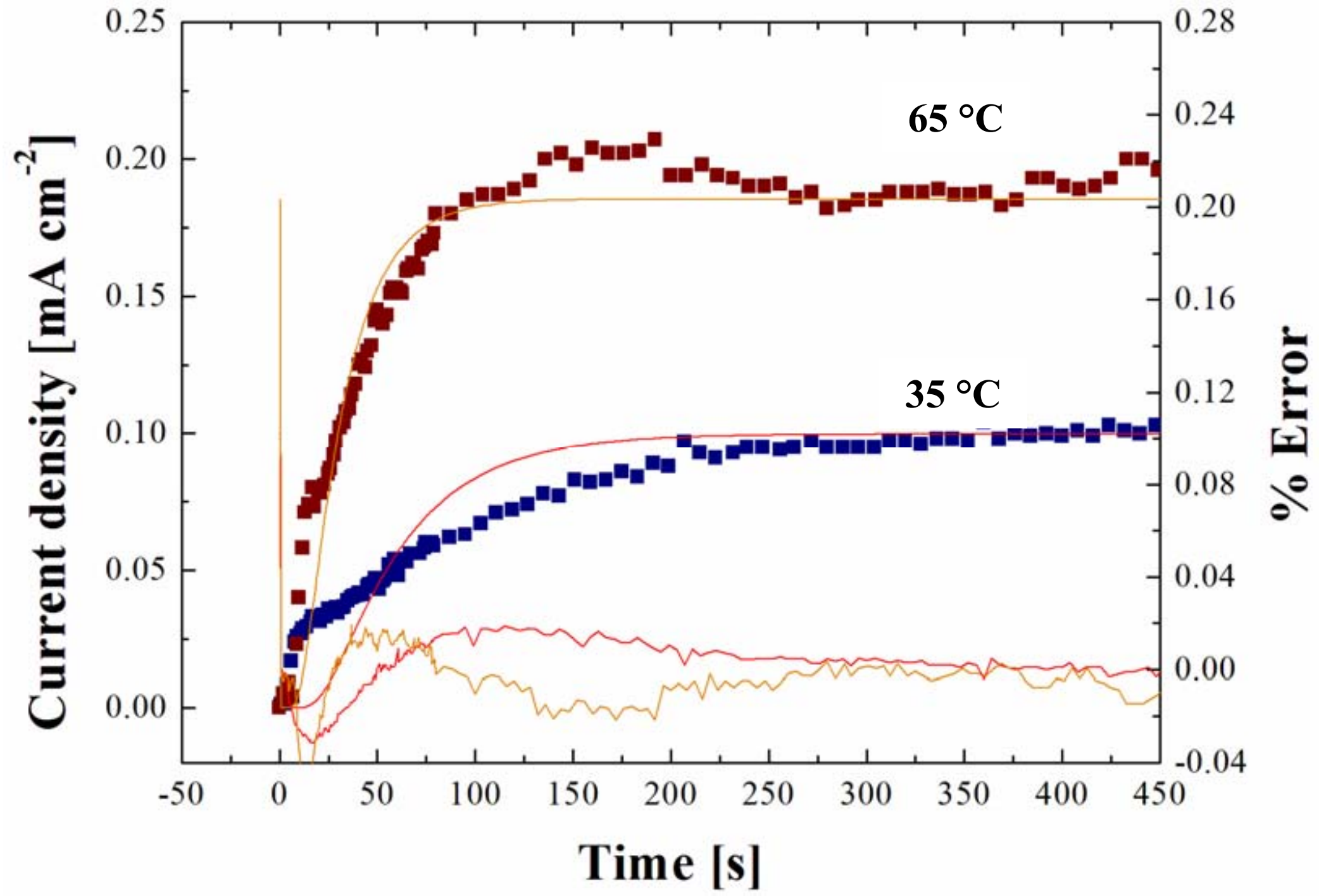

*Figure 14: Comparison of the experimental data (symbols) to equation 11 (lines) fitted with the diffusion coefficient and solubility of $H_2S$ in Nafion 112 (for 35 and 65 °C) determined from the method of least squares. The respective error between data and the fit is shown on the secondary axis.*

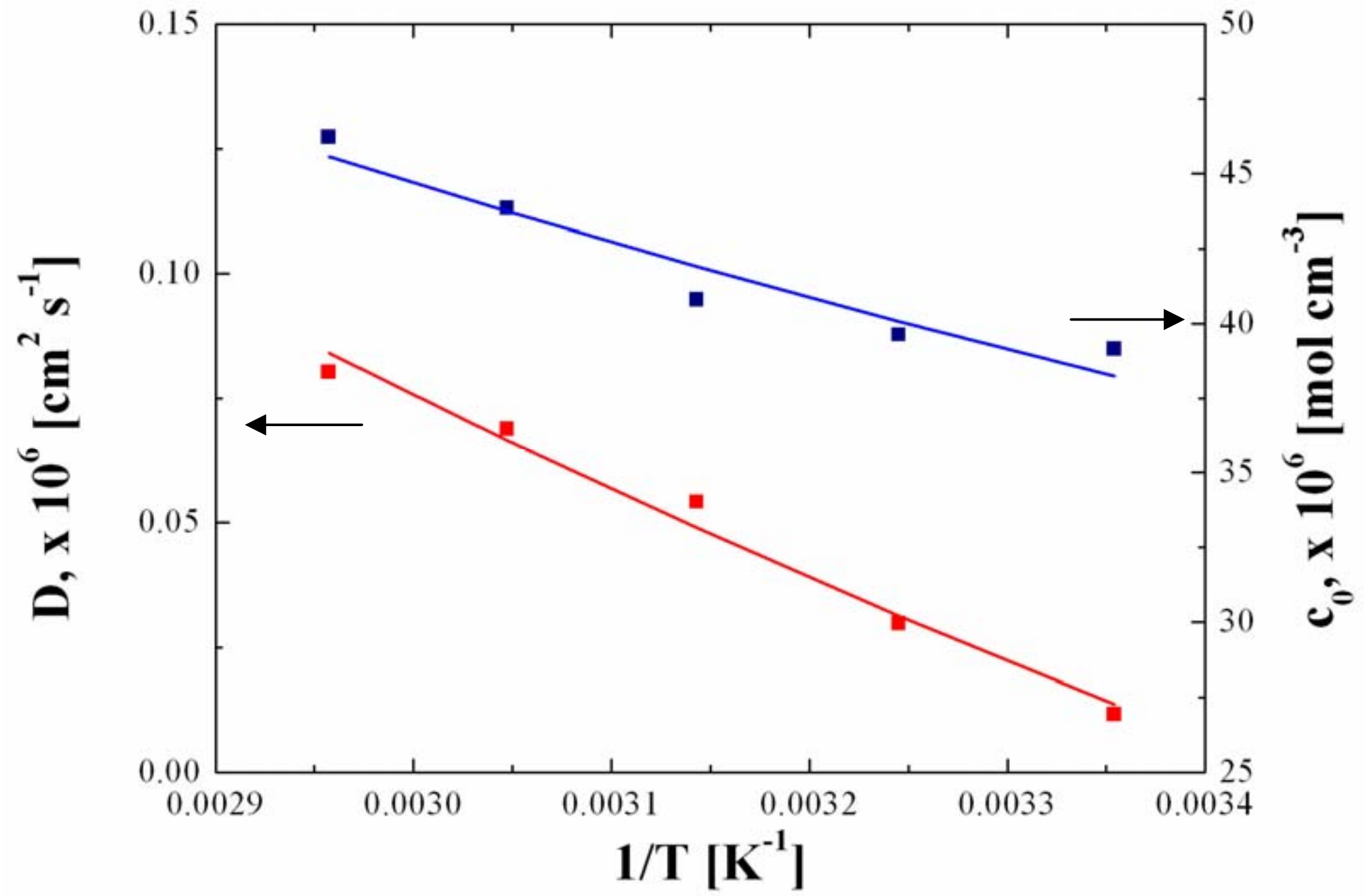

*Figure 15: Hydrogen sulfide diffusion coefficient and solubility (symbols) in Nafion 112 estimated from data obtained using the electrochemical monitoring technique. The lines are exponential fits according to equations 31 and 32. The estimated activation energy for $H_2S$ diffusion through Nafion 112 is ~8.85 kJ $mol^{-1}$ and the mixture of enthalpy is ~7.61 kJ $mol^{-1}$.*

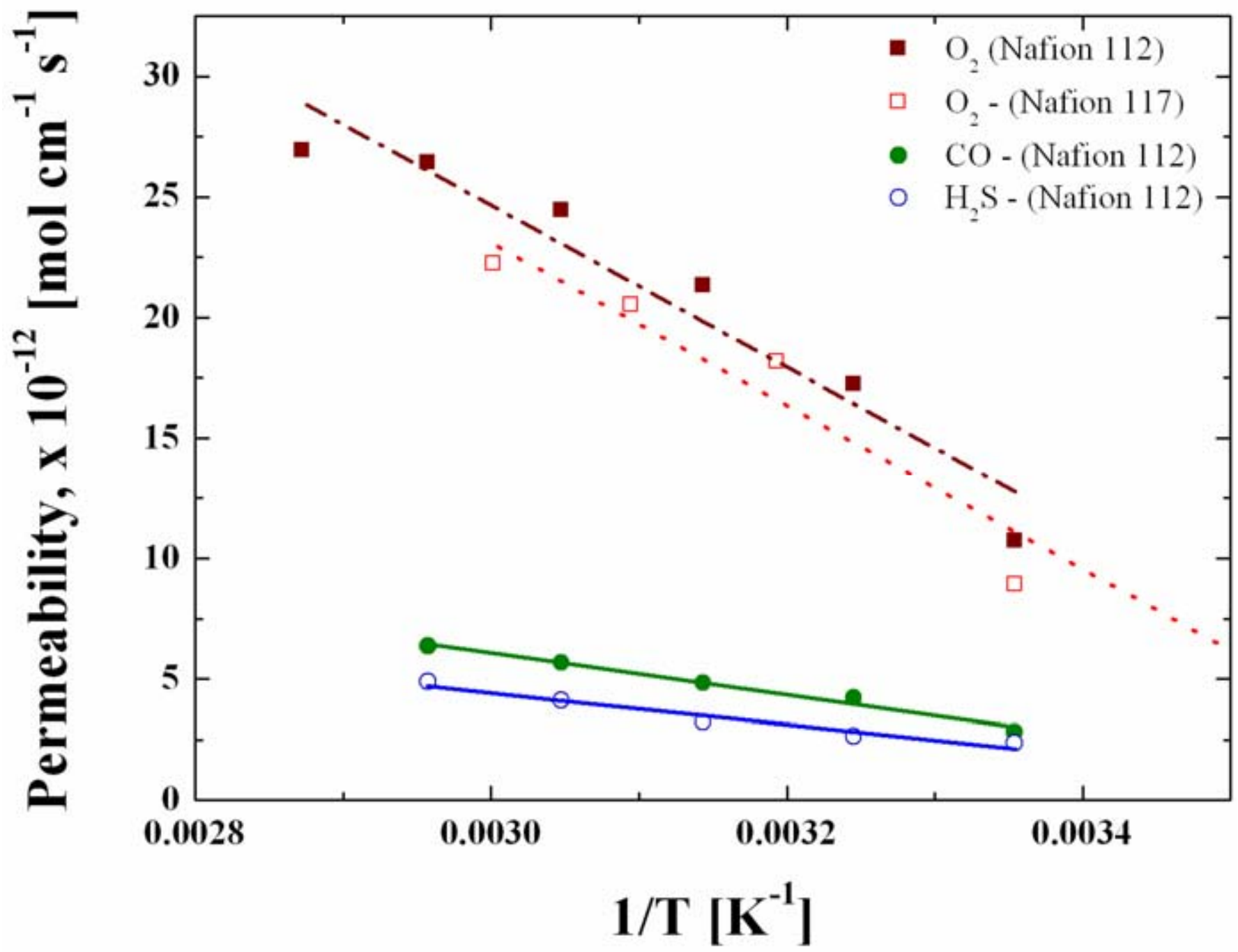

*Figure 16: Permeability of $O_2$, CO and $H_2S$ in Nafion membranes estimated using the electrochemical monitoring technique is plotted as a function of temperature. The lines represent exponential fits (i.e., shows Arrhenius dependence with temperature).*

***Figure 16*** shows the permeability of all three gases in Nafion. Since both diffusion coefficient and solubility show Arrhenius dependence with temperature, it is expected for permeability to show the same dependence. Oxygen permeability in Nafion 112 is slightly higher than in Nafion 117 even after correcting for swelling and reverse-osmosis. The increase in $O_2$ permeability with temperature is the same for both thicknesses (similar activation energies). Overall, the gas permeability in Nafion increases with temperature for all the three gases studied, as expected of gas diffusion in most polymers.

## 5. CONCLUSIONS

The electrochemical monitoring technique has been used on a Devanathan-Stachurski type diffusion cell made from a fuel cell assembly to measure the diffusion coefficient and solubility of $O_2$, CO and $H_2S$ in Nafion membranes. The diffusion cell demonstrated here has the ability to test gas crossover properties of a membrane as a function of cell temperature and gas pressure. The design can be modified easily to evaluate humidity effects. The current-time data obtained from the EMT has been used in conjunction with a Fickian model to obtain relevant transport parameters for these three gases in Nafion. Membrane swelling and reverse-gas diffusion due to water flux from the liquid cell to the gas side has been accounted for in the parameter estimation routine. The contribution from both the former and the latter were minimal. From diffusion data obtained at various temperatures of interest for PEM fuel cell operation, the activation energy for diffusion and the enthalpy of mixing have been determined for these three gases. The permeability of all three gases was observed to increase with temperature. The estimated parameters agree very well with those reported in the literature. The data reported in this paper can be used by the PEMFC modeling community and the experimental procedure can be adapted as a diagnostic tool. Since gas crossover plays an important role in both the fuel cell power output as well as its long-term durability, we recommend that this procedure be routinely used to evaluate gas transport properties of proton exchange membranes.

**Table 1: Parameters used in data analysis for Nafion 112 and 117 membranes.**

| Parameters | Value | Comments |
|---|---|---|
| $A^{Nafion\ 112}$ | 10 cm$^2$ | Measured |
| $A^{Nafion\ 117}$ | 1.76 cm$^2$ | Measured |
| $A_1$ | -66.7354 | Ref. [36] |
| $A_2$ | -169.4951 | Ref. [37] |
| $A_3$ | 263.5657 | Ref. [37] |
| $A_4$ | 159.2552 | Ref. [37] |
| $A_5$ | -25.4967 | Ref. [37] |
| $A_6$ | -3.3747 | Ref. [38] |
| $A_7$ | 0.072437 | Ref. [38] |
| $A_8$ | $-1.10765 \times 10^{-4}$ | Ref. [38] |
| $B_1$ | 87.4755 | Ref. [36] |
| $B_2$ | 0.051198 | Ref. [37] |
| $B_3$ | -0.044591 | Ref. [37] |
| $B_4$ | 0.0086462 | Ref. [37] |
| $B_5$ | -1549.159 | Ref. [38] |
| $C_1$ | 24.4526 | Ref. [36] |
| $C_2$ | 0.144237 | Ref. [38] |
| $L_0^{Nafion\ 112}$ | 0.00508 cm | Manufacturer data |
| $L_0^{Nafion\ 117}$ | 0.01778 cm | Manufacturer data |
| $k_s$ | $3.9 \times 10^{-8}$ | Ref. [44] |
| $M_m$ | 1100 g mol$^{-1}$ | Manufacturer data |
| $n_{e,\ O_2}$ | 2 | Ref. [26] |
| $n_{e,\ CO}$ | 2 | Ref. [27] |
| $n_{e,\ H_2S}$ | 6 | Ref. [28] |
| S | 0.34 | Ref. [37] |
| $s_{O_2}$ | ½ | Ref. [26] |
| $s_{CO}$ | 1 | Ref. [27] |
| $s_{H_2S}$ | 1 | Ref. [28] |
| $\bar{V}_0^{Nafion\ 112}$ | 0.0508 cm$^3$ | Measured |
| $\bar{V}_0^{Nafion\ 117}$ | 0.0313 cm$^3$ | Measured |
| $\bar{V}_m$ | 550 cm$^3$ mol$^{-1}$ | Estimated |
| $\hat{\lambda}$ | 18 | Estimated |
| $\rho_m$ | 2 g cm$^{-3}$ | Manufacturer data |
| $\gamma$ | 0.95 | Confidence |

**Table 2: Values for $O_2$ diffusion coefficient and solubility in Nafion 117 reported in the literature.**

| Source | D, x $10^6$ $cm^2$ $s^{-1}$ | $c_g$, x $10^6$ mol $cm^{-3}$ | T, °C | P, atm | Method |
|---|---|---|---|---|---|
| Haug and White [11] | 0.62 | 18.7 | 25 | 1 | EMT[b] |
| Lehtinen et al. [12] | 0.70 | 13.0 | 20 | 1 | EMT |
| | 1.9 | 9.3 | | | PST[c] |
| Parthasarathy et al. [13] | 0.74 | 26.0 | 25 | 1 | PST |
| Parthasarathy et al. [46] | 2.88 | 5.76 | 40 | 5 | |
| Ogumi et al. [9][a] | 0.24 | 7.2 | 20 | 1 | EMT |
| | 0.29 | 6.5 | 30 | | |
| | 0.44 | 5.3 | 40 | | |
| | 0.52 | 5.9 | 50 | | |
| Basura et al. [14] | 6.0 | 9.2 | 30 | 3 | PST |
| Beattie et al. [15] | 5.96 | 9.16 | 30 | 3 | PST |
| | 7.87 | 8.27 | 40 | 3 | |
| | 9.76 | 7.53 | 50 | 3 | |
| | 9.09 | 8.20 | 60 | 3 | |
| | 10.31 | 7.81 | 70 | 3 | |
| Beattie et al. [15] | 5.24 | 6.36 | 60 | 2 | PST |
| | 5.46 | 7.89 | | 2.5 | |
| | 5.48 | 9.35 | | 3 | |
| | 5.47 | 10.93 | | 3.5 | |
| | 5.71 | 11.83 | | 4 | |
| | 6.71 | 12.10 | | 4.5 | |
| | 7.07 | 12.52 | | 5.0 | |
| Chiou and Paul [40][d] | 0.0457 | 0.106 | 35 | 1 | GC[e] |
| Buchi et al. [47] | 2.6 | 4.8 | 25 | 1 | PST |
| Gode et al. [48][f] | 1.1 | 4.0 | 25 | 1 | PST |
| | 1.7 | 3.8 | 60 | 1 | |
| Parthasarathy et al. [13] | 0.995 | 9.34 | 30 | 5 | PST |
| | 2.88 | 5.76 | 40 | | |
| | 3.81 | 5.30 | 50 | | |
| | 5.23 | 4.96 | 60 | | |
| | 6.22 | 4.92 | 70 | | |
| | 8.70 | 4.43 | 80 | | |

[a] – Nafion 120
[b] – Electrochemical monitoring technique
[c] – Potential step technique
[d] – Dry Nafion 117
[e] – Gas chromatography {See Koros et al. [49] for its design and operation}
[f] – Reported values at 25 °C and 60 °C are respectively obtained at 82 % and 75% RH.

**Table 3: Diffusion coefficient, solubility and permeability of $O_2$ in Nafion 112 and Nafion 117 membranes as a function of temperature and pressure. The confidence intervals for the estimated parameters are given as well.**

**(a) Nafion 112**

| Temperature °C | Diffusion Coefficient D, x $10^6$ $cm^2\ s^{-1}$ | Confidence Interval[a] x $10^8$ $cm^2\ s^{-1}$ | Solubility $c_g$, x $10^6$ mol $cm^{-3}$ | Confidence Interval[a] x $10^6$ mol $cm^{-3}$ | Permeability $D^*c_g$, x $10^{12}$ mol $cm^{-1}s^{-1}$ |
|---|---|---|---|---|---|
| 25 | 0.1048 | 0.28 | 102.63 | 3.27 | 10.75 |
| 35 | 0.2049 | 0.40 | 84.13 | 1.87 | 17.24 |
| 45 | 0.1877 | 0.32 | 113.50 | 2.18 | 21.32 |
| 55 | 0.1715 | 0.33 | 142.72 | 3.20 | 24.47 |
| 65 | 0.1895 | 0.36 | 139.46 | 3.10 | 26.42 |
| 75 | 0.2331 | 0.77 | 115.62 | 4.49 | 26.95 |

**(b) Nafion 117**

| Temperature °C | Diffusion Coefficient D, x $10^6$ $cm^2\ s^{-1}$ | Confidence Interval[a] x $10^6$ $cm^2\ s^{-1}$ | Solubility $c_g$, x $10^6$ mol $cm^{-3}$ | Confidence Interval[a] x $10^6$ mol $cm^{-3}$ | Permeability $D^*c_g$, x $10^{12}$ mol $cm^{-1}s^{-1}$ |
|---|---|---|---|---|---|
| 10 | 0.256 | 0.023 | 23.00 | 2.24 | 5.87 |
| 25 | 0.577 | 0.055 | 15.50 | 1.38 | 8.91 |
| 40 | 1.079 | 0.170 | 16.91 | 2.23 | 18.19 |
| 50 | 1.349 | 0.173 | 15.26 | 1.98 | 20.53 |
| 60 | 1.350 | 0.154 | 16.54 | 2.58 | 22.25 |
| 70 | 1.225 | 0.291 | 17.19 | 5.13 | 20.78 |
| 80 | 1.316 | 0.457 | 14.15 | 7.49 | 18.17 |

**(c) Nafion 117**

| Pressure atm | Diffusion Coefficient D, x $10^6$ $cm^2\ s^{-1}$ | Confidence Interval[a] x $10^7$ $cm^2\ s^{-1}$ | Solubility $c_g$, x $10^6$ mol $cm^{-3}$ | Confidence Interval[a] x $10^6$ mol $cm^{-3}$ | Permeability $D^*c_g$, x $10^{12}$ mol $cm^{-1}s^{-1}$ |
|---|---|---|---|---|---|
| 1 | 0.5774 | 0.54 | 15.49 | 1.38 | 8.93 |
| 1.34 | 0.6105 | 0.40 | 22.19 | 6.12 | 13.53 |
| 1.5 | 0.6565 | 0.94 | 23.36 | 6.18 | 15.29 |
| 1.68 | 0.5744 | 0.33 | 27.65 | 6.96 | 15.86 |
| 2 | 0.6207 | 1.63 | 31.44 | 5.42 | 19.38 |
| 2.5 | 0.5956 | 0.83 | 38.87 | 2.69 | 23.11 |
| 3 | 0.6159 | 1.68 | 52.03 | 24.5 | 31.18 |

[a] – 95% confidence

**Table 4: Diffusion coefficient, solubility and permeability of CO in a Nafion 112 membrane and the confidence of the estimated parameters.**

| T °C | Diffusion Coefficient D, x $10^6$ $cm^2$ $s^{-1}$ | Confidence Interval[a] x $10^8$ $cm^2$ $s^{-1}$ | Solubility $c_g$, x $10^6$ mol $cm^{-3}$ | Confidence Interval[a] x $10^6$ mol $cm^{-3}$ | Permeability $D*c_g$, x $10^{12}$ mol $cm^{-1}s^{-1}$ |
|---|---|---|---|---|---|
| 25 | 0.1153 | 0.22 | 24.46 | 0.56 | 2.82 |
| 35 | 0.1796 | 0.39 | 23.44 | 0.60 | 4.21 |
| 45 | 0.2185 | 1.08 | 22.10 | 0.74 | 4.83 |
| 55 | 0.2595 | 1.07 | 21.85 | 1.01 | 5.67 |
| 65 | 0.3139 | 0.61 | 20.26 | 0.81 | 6.36 |

[a] – 95% confidence

**Table 5: Diffusion coefficient, solubility and permeability of $H_2S$ in a Nafion 112 membrane and the confidence of the estimated parameters.**

| T °C | Diffusion Coefficient D, x $10^6$ $cm^2$ $s^{-1}$ | Confidence Interval[a] x $10^8$ $cm^2$ $s^{-1}$ | Solubility $c_g$, x $10^6$ mol $cm^{-3}$ | Confidence Interval[a] x $10^6$ mol $cm^{-3}$ | Permeability $D*c_g$, x $10^{12}$ mol $cm^{-1}s^{-1}$ |
|---|---|---|---|---|---|
| 25 | 0.0848 | 0.34 | 26.94 | 2.84 | 2.37 |
| 35 | 0.0877 | 0.17 | 29.96 | 1.20 | 2.63 |
| 45 | 0.0948 | 0.19 | 34.02 | 0.79 | 3.22 |
| 55 | 0.1130 | 0.27 | 36.45 | 3.35 | 4.12 |
| 65 | 0.1273 | 0.41 | 38.36 | 1.50 | 4.88 |

[a] – 95% confidence

**List of Symbols**

| | |
|---|---|
| $A$ | cross-sectional area of the membrane, $cm^2$ |
| $c(x,t)$ | gas concentration at a distance, x, from the membrane and a given time, t |
| $c_g$ | solubility of gas g in Nafion, mol $cm^{-3}$ |
| $C_{kk}$ | values of the inverted approximate Hessian matrix for element k where k represents D or $c_0$ |
| $c_{g,W}$ | solubility of gas g in water, mol $cm^{-3}$ |
| $c_W$ | solubility of water in Nafion, mol $cm^{-3}$ |
| $D_g$ | diffusion coefficient of gas g, $cm^2$ $s^{-1}$ |
| $D_{W,F}$ | Fickian diffusion coefficient for water in Nafion, $cm^2$ $s^{-1}$ |
| $E_a$ | activation energy, kJ $mol^{-1}$ |
| $F$ | Faraday's constant, 96487 C $equiv^{-1}$ |
| $i(t)$ | current, A |
| $i_\infty$ | limiting current, A |
| $K_{O_2}$ | Henry's law constant for $O_2$ |
| $k_s$ | first ionization constant for $H_2S$ in water |
| $L$ | thickness of membrane, cm |
| $M_m$ | molecular weight of the membrane, g $mol^{-1}$ |
| $m$ | number of variables |
| $n$ | number of data points taken in each trial |
| $n_e$ | number of electrons transferred |
| $P_{g,W}$ | permeability of gas g due to water flux from the liquid side to gas side, mol $cm^{-1}$ $s^{-1}$ |
| $P_k$ | $k^{th}$ parameter |
| $\hat{P}_k$ | estimate of the $k^{th}$ parameter |
| $R$ | universal gas constant, 8.314 J $mol^{-1}$ $K^{-1}$ |
| $s_g$ | stoichiometric coefficient of gas g |
| $s_{\hat{P}_k}$ | standard deviation |
| $x$ | distance from the catalyst-layer, cm |
| $t_\gamma$ | value of the t distribution |
| $T$ | temperature, K (or °C) |
| $t$ | time, s |
| $\bar{V}_0$ | initial volume of the membrane, $cm^3$ |
| $\bar{V}_m$ | partial molar volume of the membrane, $cm^3$ $mol^{-1}$ |
| $X_{O_2,W}$ | mole fraction solubility of $O_2$ in pure liquid water at 101 kPa |
| $X_{CO,W}$ | solubility of CO in pure liquid water, nL $L^{-1}$ |
| $\lambda$ | moles of water per mole of sulfonic acid sites (water content) |
| $\hat{\lambda}$ | average membrane water content |
| $\rho_m$ | density of the membrane, g $cm^{-3}$ |
| $\gamma$ | confidence |

**ACKNOWLEDGEMENTS**

The authors gratefully acknowledge the support from the National Reconnaissance Office for *Hybrid Advanced Power Sources* under grant # NRO-00-C-1034 and the National Science Foundation for funding Saahir Khan under the *Research Experience for Undergraduates (REU)* Program.